\documentclass[fleqn,usenatbib]{mnras}

\usepackage{mathptmx}

\usepackage[T1]{fontenc}

\usepackage{graphicx}	
\usepackage{amsmath}	
\usepackage{amssymb}	
\usepackage{natbib}
\usepackage{bm}

\usepackage{color}
\usepackage{ulem}

\def\Vec#1{\mathbfit #1}

\def\P#1#2{\dfrac{\partial #1}{\partial #2}}

\title[Magnetic mountains with multipole magnetic fields]
{Magnetically confined mountains on accreting neutron stars with multipole magnetic fields}

\author[K. Fujisawa et al.]{
Kotaro Fujisawa$^{1}$\thanks{E-mail: fujisawa@resceu.s.u-tokyo.ac.jp},
Shota Kisaka$^{2}$ \& Yasufumi Kojima$^{2}$ \\
$^{1}$Department of Physics, Graduate School of Science, the University of Tokyo, Bunkyo-ku, Tokyo 113-0033, Japan, \\
$^{2}$Department of Physics, Graduate School of Advanced Science and Engineering, Hiroshima University,  Higashi-Hiroshima, Hiroshima 739-8526, Japan\\
}

\date{Accepted 2022 September 8. Received 2022 September 7; in original form 2022 June 24   }

\pubyear{2022}

\begin{document}
\label{firstpage}
\pagerange{\pageref{firstpage}--\pageref{lastpage}}
\maketitle

 \begin{abstract}
Magnetically confined mountains on accreting neutron stars are candidates for producing continuous gravitational waves. We formulate a magnetically confined mountain on a neutron star with strong multipole magnetic fields and obtain some sequences of numerical solutions. We find that the mass ellipticity of the mountain increases by one order of magnitude if the neutron star has strong multipole magnetic fields. As matter accretes on to the magnetic pole, the size of the mountain increases and the magnetic fields are buried. If the neutron star has a dipole magnetic field, the dipole magnetic field is buried and transformed into multipole components. By contrast, if the neutron star has both dipole and strong multipole magnetic fields, the multipole magnetic fields are buried and transformed into a negative dipole component. We also calculate magnetically confined mountains with toroidal magnetic fields and find that the ellipticity becomes slightly smaller when the mountain has toroidal magnetic fields. If the multipole magnetic fields are buried, they sustain the intense toroidal magnetic field near the stellar surface, and the ratio of the toroidal magnetic field to the poloidal magnetic field is close to 100. The hidden strong toroidal magnetic fields are sustained by the buried multipole magnetic fields. 
\end{abstract}




\section{Introduction}

A gravitational wave is a new tool for understanding various astronomical phenomena and objects. 
Continuous gravitational waves, which are monochromatic and have constant amplitude, are thought to be produced by fast-spinning objects, such as neutron stars. Suppose such a neutron star is non-axisymmetrically distorted around its rotational axis or has a mountain on its surface. The mass quadrupole moment of the neutron star depends on the rotational phase, and the continuous gravitational wave is emitted. Although continuous gravitational waves from neutron stars have not yet been detected, the upper limit of the mass ellipticity is becoming more strictly limited (\citealp{2014ApJ...785..119A, 2015PhRvD..91f2008A, 2017PhRvD..96l2006A, LIGO_2017ApJ, 2017ApJ...847...47A, 2018PhRvD..97l9903A,  2019PhRvD.100b4004A, 2019PhRvD.100l2002A,  2019ApJ...879...10A, 2019ApJ...882...73A, 2020ApJ...899..170A, 2020ApJ...902L..21A,  LIGO_2021PRD, 2021PhRvD.104j9903A, 2021ApJ...913L..27A, 2021ApJ...921...80A, 2021ApJ...922...71A,  LIGO_2021arXiv_a, 2022PhRvD.105h2005A, LIGO_2021arXiv_b, 2022arXiv220404523T}).

There are many possible mechanisms for inducing non-axisymmetric deformation and mountains (\citealp{Haskell_et_al_2015}). The elastic strain of the crust (\citealp{Ushomirsky_et_al_2000, Haskell_et_al_2006}) is one of the mechanisms. The maximum size of the mountain that an elastic crust can support is determined by the maximum strain. Recently, \citet{Gittins_Andersson_Jones_2021} pointed out that there were issues relating to boundary conditions in previous works that must be satisfied for realistic neutron stars. They showed that the maximum size of the mountain is much smaller than previously suggested. They also showed that the maximum size of the mountain depends on the evolution history, such as solidification of the crust and spin-down of the neutron star. The size of the mountain is further suppressed by accounting for general relativistic gravity (\citealp{Gittins_Andersson_2021}). It could be difficult for an elastic crust to sustain a mountain on an isolated neutron star. In a binary system, on the other hand, a mountain may have other physical origins that caused accretion. 

Magnetically confined mountains are one of the possible mechanisms in an accreting system (\citealp{Melatos_Phinney_2001, Payne_Melatos_2004, Priymak_et_al_2011, Mukherjee_2017, Suvrov_Melatos_2019, Singh_et_al_2020}). On the accreting neutron star surface, matter from the companion star is transferred along magnetic field lines and accreted on to the magnetic poles, becoming a magnetically confined mountain supported by magnetic force. In the case of a neutron star having a dipole magnetic field, the matter accretes on to both the northern and southern magnetic poles, and two mountains are formed symmetrically near the magnetic poles (\citealp{Payne_Melatos_2004, Priymak_et_al_2011}). However, in reality, high-quality observations have revealed different configurations, suggesting the star's magnetic field is not a simple dipole. 

Recently, the Neutron star Interior Composition ExploreR (NICER) obtained high-quality pulse profiles of millisecond pulsars PSRs J0030+0451 (\citealp{Bilous_et_al_2019,Miller_et_al_2019,Riley_et_al_2019}) and J0740+6620 (\citealp{2021ApJ...918L..27R,2021ApJ...918L..28M}). According to the observations, hot spots on the surface are not symmetrically located with one in the northern and one in the southern hemisphere, but rather with multiple hot spots in the same hemisphere in asymmetrical positions. This indicates that the magnetic field at the surface of the star is not a simple dipole but a more complex multipole magnetic field (\citealp{Bilous_et_al_2019,2020ApJ...893L..38C,2021ApJ...907...63K,2021ApJ...918L..27R}). 

Multipole magnetic fields are related to magnetically confined mountains on accreting neutron stars. \citet{Priymak_et_al_2011} calculated magnetically confined mountains of a neutron star with a dipole magnetic field. As accretion on to the magnetic poles progresses, the magnetic field lines become distorted. The dipole field is buried by the accreted matter, and the magnetic dipole moment is transformed into higher-order multipole moments. The magnetic burial due to the accretion induces multipole magnetic fields. \citet{Suvorov_Melatos_2020} considered an equatorially asymmetric accretion flow and calculated the resulting magnetically confined mountains. The asymmetric accretion also forms the higher-order magnetic fields, which contain both odd- and even-order magnetic moments. The latter are induced by equatorially asymmetric accretion. However, previous works have assumed that the neutron star's magnetic field is a dipole. Thus, the effects of strong multipole magnetic fields on magnetically confined mountains remain unclear. The toroidal magnetic field, which is important for stabilizing the magnetic field configuration, has also been ignored by  previous studies.

We investigate how the magnetic fields are buried and what kinds of mountains are formed when a neutron star has a dipole magnetic field and strong multipole magnetic fields. We consider the case where the strength of the multipole magnetic field is larger than that of the dipole field. We also include the toroidal magnetic fields. Since formulations and numerical methods in previous studies assumed a neutron star with only a dipole magnetic field, we cannot directly apply these to a system with strong multipole and toroidal magnetic fields. Therefore, we develop a new numerical method for calculating magnetically confined mountains. We formulate the problem based on magnetized stars with twisted-torus magnetic fields (\citealp{Tomimura_Eriguchi_2005, Yoshida_Eriguchi_2006, Yoshida_Yoshida_Eriguchi_2006, Lander_Jones_2009, Fujisawa_Yoshida_Eriguchi_2012, Fujisawa_Kisaka_2014}). Although the numerical method for magnetized stars is different from those used in previous studies for magnetically confined mountains, we have confirmed that our new method is suitable for this application. This paper is organized as follows. In Section 2, we describe our new formulation and numerical method for obtaining magnetically confined mountains with strong multipole magnetic fields. Numerical results are given in Section 3. Section 4 provides discussion and Section 5 concludes. 

\section{Formulation}

\subsection{Basic equations and formulation}

We assume that the system is stationary and axisymmetric around the magnetic axis and ignore the star's viscosity, meridional flow, and rotation. We also assume that the gravity is Newtonian, the magnetic mountain is barotropic, and the magnetic fields have both poloidal and toroidal components. Under these assumptions, the Euler equation to describe the magnetic mountain becomes
\begin{align}
    \frac{1}{\rho} \nabla p = - \nabla \phi  
    + \frac{1}{\rho} \left( \frac{\Vec{j}}{c} \times \Vec{B} \right),
    \label{Eq:Euler}
\end{align}
where $\rho$, $p$, $\phi$, $c$, $\Vec{j}$, and $\Vec{B}$ are, respectively, the mass density, pressure, gravitational potential of the star, speed of light, electric current density, and magnetic field. Since the mountain is barotropic, $p$ is a function of $\rho$. We ignore the self-gravity of the mountain, but consider the external gravitational force from the neutron star, whose potential is given by
\begin{align}
    \phi(r,\theta) = - \frac{G M_{NS}}{r},
\end{align}
where $G$ and $M_{NS}$ are the gravitational constant and the mass of the neutron star, respectively. Spherical polar coordinates $(r, \theta, \varphi)$ are employed in this paper. We introduce the magnetic flux function $\Psi$ as follows: 
\begin{align}
B_r = \frac{1}{r^2 \sin \theta}\P{\Psi}{\theta}, \ B_\theta = 
- \frac{1}{r \sin\theta}\P{\Psi}{r}.
\end{align}
Using $\Psi$, the $\varphi$ component of Maxwell's equation is
\begin{align}
  \Delta^* \Psi \equiv  \frac{\partial^2 \Psi}{\partial r^2} + 
\frac{\sin \theta}{r^2} \frac{\partial}{\partial \theta} \left( 
\frac{1}{\sin \theta} \frac{\partial \Psi}{\partial \theta} \right)
= - 4 \pi r \sin \theta \frac{j_\varphi}{c}.
\label{Eq:GS}
\end{align}
This elliptic-type equation \eqref{Eq:GS} is called the Grad-Shafranov (GS) equation. The source term  $j_\varphi$ is determined by the Euler equation (equation \ref{Eq:Euler}). From the axisymmetry condition, the $\varphi$ component of the curl of the Euler equation should vanish, and  the toroidal magnetic field is constrained as
\begin{align}
\left( \frac{1}{\rho} \frac{\Vec{j}}{c} \times \Vec{B} \right)_\varphi = 0  \Rightarrow   B_\varphi = \frac{S(\Psi)}{r \sin \theta},
\end{align}
where $S(\Psi)$ is an arbitrary function of $\Psi$. We assume the polytropic type equation of state as
\begin{align}
    p = K \rho^{\Gamma},
\end{align}
where $K$ and $\Gamma$ are constants, and the Lorentz force should be irrotational as 
\begin{align}
    \nabla \times \left( \frac{1}{\rho} \frac{\Vec{j}}{c} \times \Vec{B} \right) =
    \nabla \times \nabla \phi + 
    \nabla \times \left( \frac{1}{\rho} \nabla p \right) = 0.
    \label{Eq:rot_F}
\end{align}
The Lorentz force should be a gradient of a scalar function $F$ as
\begin{align}
    \frac{1}{\rho} \left( \frac{\Vec{j}}{c} \times \Vec{B} \right) = \nabla F(\Psi),
\label{Eq:nabla_F1}
\end{align}
where $F(\Psi)$ is an arbitrary function of $\Psi$. Using the function $F(\Psi)$, we analytically integrate equation \eqref{Eq:Euler}, and obtain the first integral as
\begin{align}
            \frac{1}{\rho} \nabla p &= - \nabla \phi + \nabla F(\Psi) \nonumber \\
\Rightarrow    \int \frac{dp}{\rho} &= K \frac{\Gamma}{\Gamma - 1} \rho^{\Gamma - 1} 
= - \phi + F(\Psi) + C, 
\label{Eq:first_int}
\end{align}
where $C$ is an integral constant. Using these arbitrary functions $F(\Psi)$ and $S(\Psi)$, the electric current density $j_\varphi$ is described as
\begin{align}
    \frac{j_\varphi}{c} = \frac{S(\Psi) S'(\Psi)}{4\pi r\sin \theta} + \rho r \sin \theta F'(\Psi),
    \label{Eq:GSsource}
\end{align}
where $S'(\Psi)$ and $F'(\Psi)$ are, respectively, derivatives of $S(\Psi)$ and $F(\Psi)$ with respect to $\Psi$. The mass density and pressure are calculated from equation \eqref{Eq:first_int} as
\begin{align}
    \rho = \left( \frac{ - \phi + C + F(\Psi)}{K \frac{\Gamma}{\Gamma -1 } }\right)^{\frac{1}{\Gamma - 1}}, \hspace{10pt}
    p = K  \left( \frac{ - \phi + C  + F(\Psi)}{K \frac{\Gamma}{\Gamma -1 } }\right)^{\frac{\Gamma}{\Gamma - 1}}.
    \label{Eq:rho_p}
\end{align}
We obtain one equilibrium solution of a mountain by fixing two functional forms $F(\Psi)$ and $S(\Psi)$, and solving the equation \eqref{Eq:GS} with the source in equation \eqref{Eq:GSsource} and appropriate boundary conditions.

\subsection{Boundary conditions and functional forms}

We describe how to fix the functional forms and impose the boundary conditions for the GS equation. We first consider that, in this context, a magnetic field flux function $\Psi$ consists of two components, the neutron star's magnetic field $\Psi_{NS}$ and the mountain's magnetic field $\Psi_{M}$:
\begin{align}
    \Psi = \Psi_{M} + \Psi_{NS}.
    \label{Eq:Psi_M_Psi_NS}
\end{align}
In this paper we consider the effect of multipole magnetic fields and treat the neutron star's magnetic field $\Psi_{NS}$ as external fields in a vacuum, for which $\Delta^* \Psi_{NS} = 0$. The magnetic flux function $\Psi_{NS}$ for a combination of dipole, quadrupole, and octupole magnetic fields is described as follows:
\begin{align}
    \Psi_{NS}(r,\theta) &= \frac{B_d r_s^3}{2} \frac{\sin^2 \theta}{r}
    + \frac{B_q r_s^4}{2}\frac{\sin^2\theta \cos\theta}{r^2} \\ \nonumber 
    &+ \frac{B_o r_s^5}{8}\frac{\sin^2 \theta (5 \cos^2 \theta - 1) }{r^3},
\end{align}
where $r_s$ is the stellar radius, and $B_d$, $B_q$, and $B_o$ are the strength of the dipole, quadrupole, and octupole magnetic fields, respectively, at the north pole.  

The mountain component of the magnetic flux $\Psi_M$ is obtained by solving the GS equation as,
\begin{align}
  \Delta^* \Psi_{M} = -4\pi r \sin \theta\frac{j_\varphi}{c} = 
  - S(\Psi) S'(\Psi) - 4\pi \rho r^2 \sin^2 \theta F'(\Psi).
\end{align}
This equation is a nonlinear elliptic type, and we should impose boundary conditions. We fix the magnetic flux function on the stellar surface ($r=r_s$) as 
\begin{align}
\Psi(r_s,\theta) = \Psi_{NS}(r_s, \theta) \Rightarrow \Psi_{M}(r_s,\theta) = 0.
\end{align}
We impose the outer boundary condition as
\begin{align}
    \Psi \Rightarrow 0 \,\, (r\Rightarrow \infty).
\end{align}
On the symmetric axis $\theta = 0$ and $\theta = \pi$, we impose the following regularity condition:
\begin{align}
    \P{\Psi}{\theta} = 0.
\end{align}
Previous studies (\citealp{Payne_Melatos_2004, Priymak_et_al_2011}) imposed similar boundary conditions and solved the GS equation (equation~\ref{Eq:GS}) directly. By contrast, we derive the integral form of the GS equation imposing the boundary conditions. Since the mountain exists within a finite region, these boundary conditions are satisfied using a Green's function and the integral form of the GS equation (\citealp{Fujisawa_Kisaka_2014}) as
\begin{align}
 \Psi_{M} \sin \varphi
  = \frac{1}{c} r \sin \theta \int \frac{j_\varphi \sin \varphi'}{|\Vec{r} - \Vec{r}'|} dV'+ 
  \Psi_{h} (r,\theta) \sin \varphi,
  \label{Eq:PsiM}
\end{align}
  where $\Psi_{h}$ is the homogeneous (general) solution to the GS equation as follows:
  \begin{align}
      \Psi_{h}(r,\theta) = \sum_{n=0}^{n_{\max}} a_n \frac{r_s^{n+2}}{r^{n}} \sin \theta P_{n}^1 (\cos \theta),
  \end{align}
  where $P_{n}^1 (\cos \theta)$ are associated Legendre polynomials and the coefficients $a_n$ are determined to satisfy the condition at the surface $\Psi_{M}(r_s, \theta) = 0$ $(\Psi(r_s, \theta)\ = \Psi_{NS})$(see \citealp{Fujisawa_Kisaka_2014}). Note that both outer and regularity boundary conditions are automatically satisfied by using equation \eqref{Eq:PsiM}.

In order to solve the GS equation, we should fix two functional forms $S(\Psi)$ and $F(\Psi)$. The latter, which is a potential for the Lorentz force, is related to the accretion's mass distribution. Among the almost arbitrary functional forms of $F$, it should be determined from physical considerations. \citet{Payne_Melatos_2004} considered the mass-flux inside the magnetic field lines (see details in Appendix \ref{App:A}) and modelled a fitting formula for $F(\Psi)$ as
\begin{align}
    F(\Psi) &= \exp(-\Tilde{\Psi})(0.027 \Tilde{\Psi}^4 - 0.13 \Tilde{\Psi}^3 
    + 0.21 \Tilde{\Psi}^2 - 0.021\Tilde{\Psi} + 0.1333)  \nonumber \\
\Tilde{\Psi} &= \Psi / (a \Psi_c),
\label{Eq:PM_Psi}
\end{align}
where $\Psi_c$ is a maximum magnetic flux function on the surface and $a$ is a positive constant. However, the functional form in equation \eqref{Eq:PM_Psi} has a discontinuity at a magnetic pole, $\Psi = 0$. The discontinuity at $\Psi = 0$ is inconvenient because we must extend the negative region of the magnetic flux function to calculate models with a multipole field. We therefore propose the following functional form, which behaves similarly to equation \eqref{Eq:PM_Psi} and is everywhere continuous:
\begin{align}
    F(\Psi) &= \frac{F_0}{\cosh(\Tilde{\Psi})}  \nonumber \\
\Tilde{\Psi} &= \Psi / (a \Psi_c),
\label{Eq:F_Psi}
\end{align}
where $F_0$ is a constant and we set $a=3$ in this paper. This functional form reproduces the numerical results of previous works (Appendix \ref{App:A}). We also use the functional form in equation \eqref{Eq:F_Psi} to calculate mountains with multipole magnetic fields, although the accretion process with multipole magnetic fields is in general complicated  (\citealt{Das_et_al_2022}) and the above formula is not unique. There are many possible functional forms for $F(\Psi)$ (\citealt{Mukherjee_2017, Singh_et_al_2020}).

The functional form of $S$ determines the distribution of the toroidal magnetic fields. To avoid the discontinuity of the toroidal magnetic fields, we fix the following functional form:
\begin{align}
    S(\Psi) =
\begin{cases}
S_0 \Psi (\Psi / \Psi_{\max} - 1 ) &   \Psi \geq \Psi_{\max} \\
S_0 \Psi (\Psi / \Psi_{\min} - 1 ) &   \Psi \leq \Psi_{\min} \\
0 & \Psi_{\min} < \Psi < \Psi_{\max},
\end{cases}
\label{eq:S_Psi}
\end{align}
where $\Psi_{\min}$ and $\Psi_{\max}$ are the minimum and maximum values of $\Psi$ in the vacuum region, respectively. This functional form is usually used for the magnetized equilibrium state (\citealt{Tomimura_Eriguchi_2005}, \citealt{Fujisawa_Kisaka_2014}). The toroidal magnetic field is confined within the closed-loop inside the mountain using this functional form. 

\subsection{Global physical quantities}

We introduce some global physical quantities to characterize the numerical solutions. The absolute value of the ellipticity determines the amplitude of the continuous gravitational wave from a spinning neutron star. The ellipticity of the mountain $\varepsilon$ is given \citep[e.g.,][]{Mastrano_et_al_2011} as
\begin{align}
\varepsilon = \frac{I_{zz} - I_{xx}}{I_0}
    &= \frac{\pi}{I_0} \int_0^\pi \int_0^\infty 
    \rho(r,\theta) r^4 \left(1 - 3\cos^2 \theta \right) \sin \theta dr d\theta, 
\end{align}
where $I_0$ is the fiducial moment of the inertia of the spherical neutron star, and $I_{ii}$ is the moment of the inertia around the $i$-axis, calculated as
\begin{align}
    I_{ii} = \int \rho(r,\theta) (r^2 - x_{i}^2) dV.
\end{align}
We set $I_0 = 1.6 \times 10^{45} \mathrm{g~cm^2}$ as a typical value for a neutron star.  

Following \citet{Priymak_et_al_2011}, we calculate the $\ell$-th magnetic multipole moment $\mu_\ell(r)$ as
\begin{align}
    \mu_\ell (r) = \frac{\ell(2\ell + 1) }{2(\ell +1)} r^\ell
    \int_0^\pi \Psi(r,\theta) P_\ell^1 (\cos \theta') d\theta. 
\end{align}
To compare models with different $\ell$, it is convenient to introduce normalized moments $\tilde{\mu_\ell}$ (\citealt{Suvorov_Melatos_2020}) defined as
\begin{align}
    \tilde{\mu}_\ell(r) = \frac{2(\ell+1)}{r^{\ell - 1} \ell (2\ell+1)}\mu_\ell(r).
\end{align}
We calculate the normalized moments at the stellar surface $r=r_s$ and the outer boundary of the computational domain $r=r_{\mathrm{out}}$. Since we fix the magnetic field at the surface, the values of $\tilde{\mu}_\ell(r_s)$ are also fixed. If the $\ell$-th magnetic moment at the outer boundary vanishes ($\tilde{\mu}_\ell(r_{\mathrm{out}}) = 0$), the $\ell$-th magnetic field is buried completely. 

Since our goal in this paper is to study the effect of strong multipole fields on the structures of the mountain, for simplicity, we treat the neutron star's surface as a rigid solid where the mountain does not sink into the crust. Note that a high-density mountain might sink, and that ellipticity decreases by a factor of $2-3$ (\citealt{Wette_et_al_2010, Suvrov_Melatos_2019}).

\subsection{Numerical settings}

\begin{table}
    \centering
    \begin{tabular}{ccccc}
Model & $B_d/B_0$ & $B_q / B_0$ & $B_o / B_0$ & $B_\varphi$\\   
\hline
d1    &  1   & 0 & 0  & 0 \\
q10 &   1 &  10  & 0 & 0\\
o10 &   1 &  0 & 10  & 0 \\
q10t & 1 & 10 & 0 & $\neq 0$\\
o10t & 1 &      0 & 10 & $\neq 0$\\
    \end{tabular}
    \caption{Model parameters of a neutron star's magnetic field. $B_0 = 3.0\times 10^{12}$G is a fiducial value of the magnetic field. Model d1 has a dipole magnetic field. Models q10 and o10 have strong quadrupole and octupole magnetic fields, respectively. Models q10t and o10t each have a non-zero toroidal magnetic field.}
    \label{tab:param}
\end{table}

Table \ref{tab:param} shows the parameter settings of the numerical models. We set the fiducial value of the magnetic field as $B_0 = 3.0 \times 10^{12}$G. For example, model d1 has a dipole magnetic field $B_d = 3.0 \times 10^{12}$G at the north pole. The structure and ellipticity of the mountain depend on the equation of state. Following previous works (\citealt{Priymak_et_al_2011}), we use a realistic accreted matter model as $K = 6.18 \times 10^{15}$ (in cgs unit) and $\Gamma = 1.18$ in equation~\eqref{Eq:rho_p}. We consider locally strong multipole magnetic fields in models q10 and o10 in Table~\ref{tab:param}. These models have both dipole and mulutipole magnetic fields. The quadrupole magnetic field $B_q$ (model q10) or octupole magnetic field $B_o$ (model o10) is ten times larger than the dipole magnetic field $B_d$ at the north pole. A model with $t$ has a non-zero toroidal magnetic field. We set the mass of the neutron star $M_{NS} = 1.4~\mathrm{M}_\odot$. The inner boundary $r_{\mathrm{in}}$ of the computational domain is set at the stellar surface, $r_{s} = 12\mathrm{km}$, while the outer boundary $r_{\mathrm{out}}$ is set above of the peak of the highest mountain (Appendix~\ref{App:A}). The number of grid points in the $r$ direction within $r \in [r_{\mathrm{in}}, r_{\mathrm{out}}]$ and that in the $\theta$ direction within $\theta \in [0,\pi]$ are defined as $N_r$ and $N_\theta$, respectively. Following previous works, we use $N_r = 257$ and $N_\theta = 513$, which spport convergence of the numerical results.

\section{Numerical results}

\subsection{Dipole magnetic field burial}

\begin{figure*}
    \centering

   \includegraphics[width=0.49\textwidth]{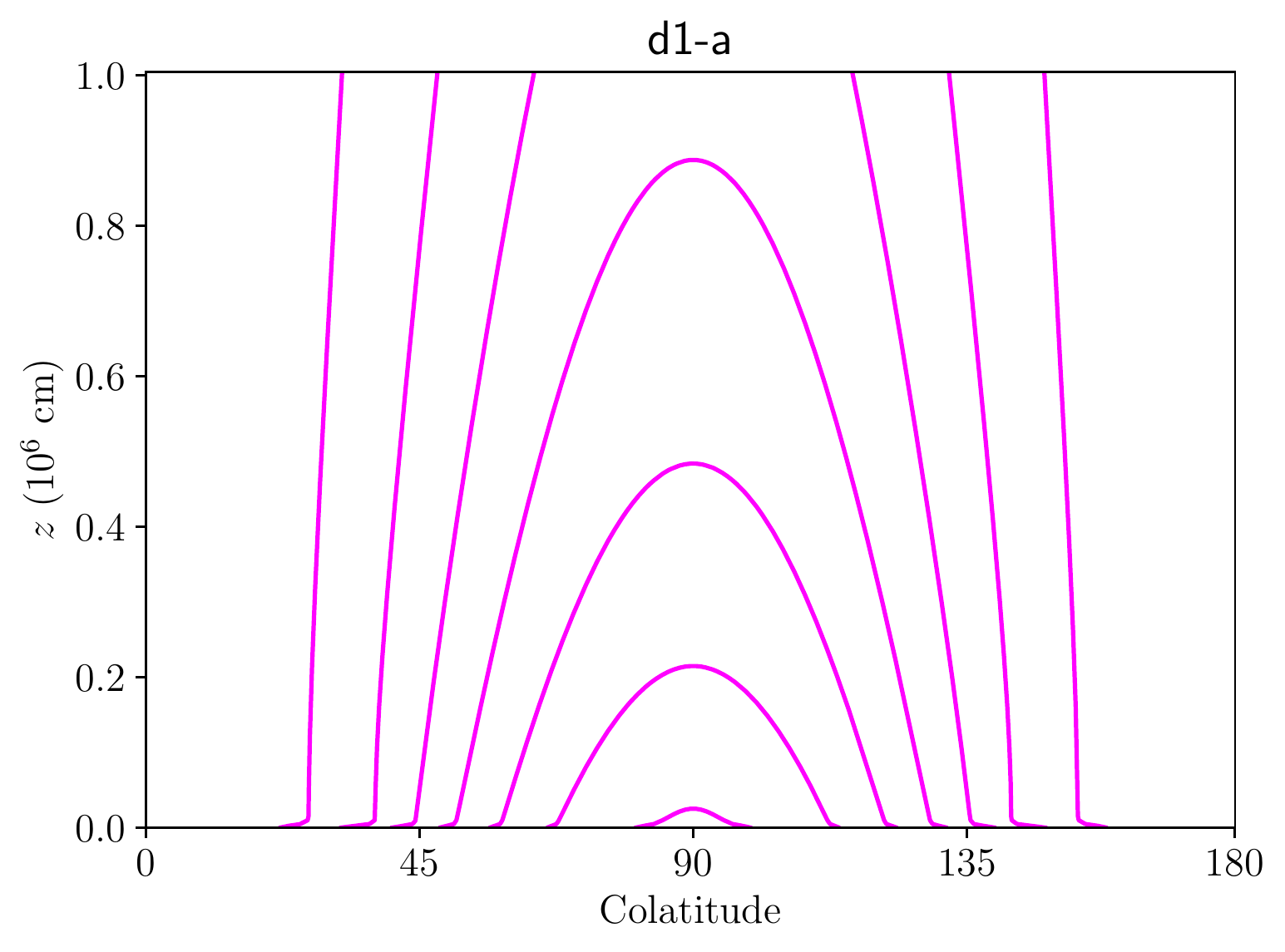}
   \includegraphics[width=0.49\textwidth]{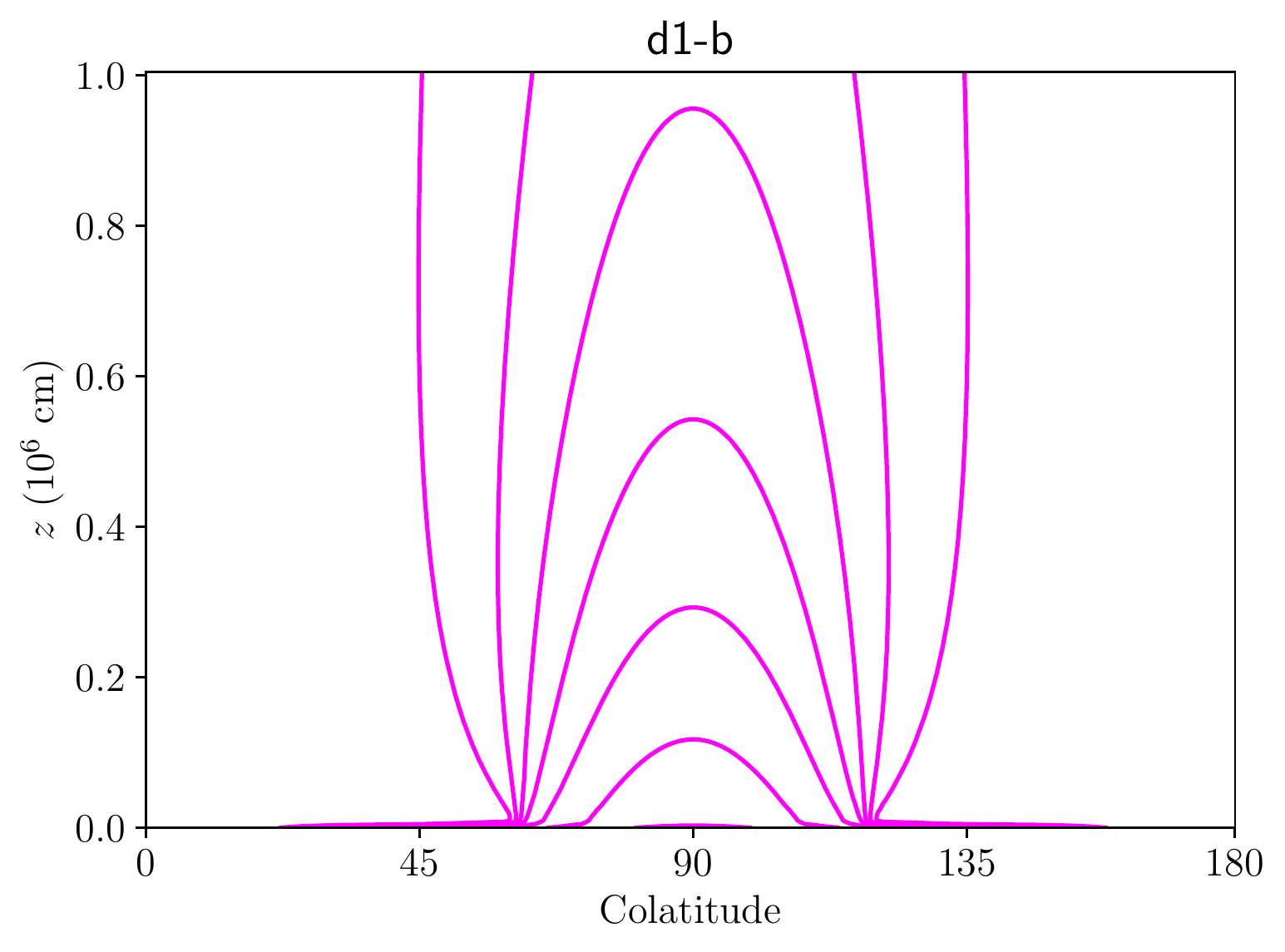}

   \includegraphics[width=0.49\textwidth]{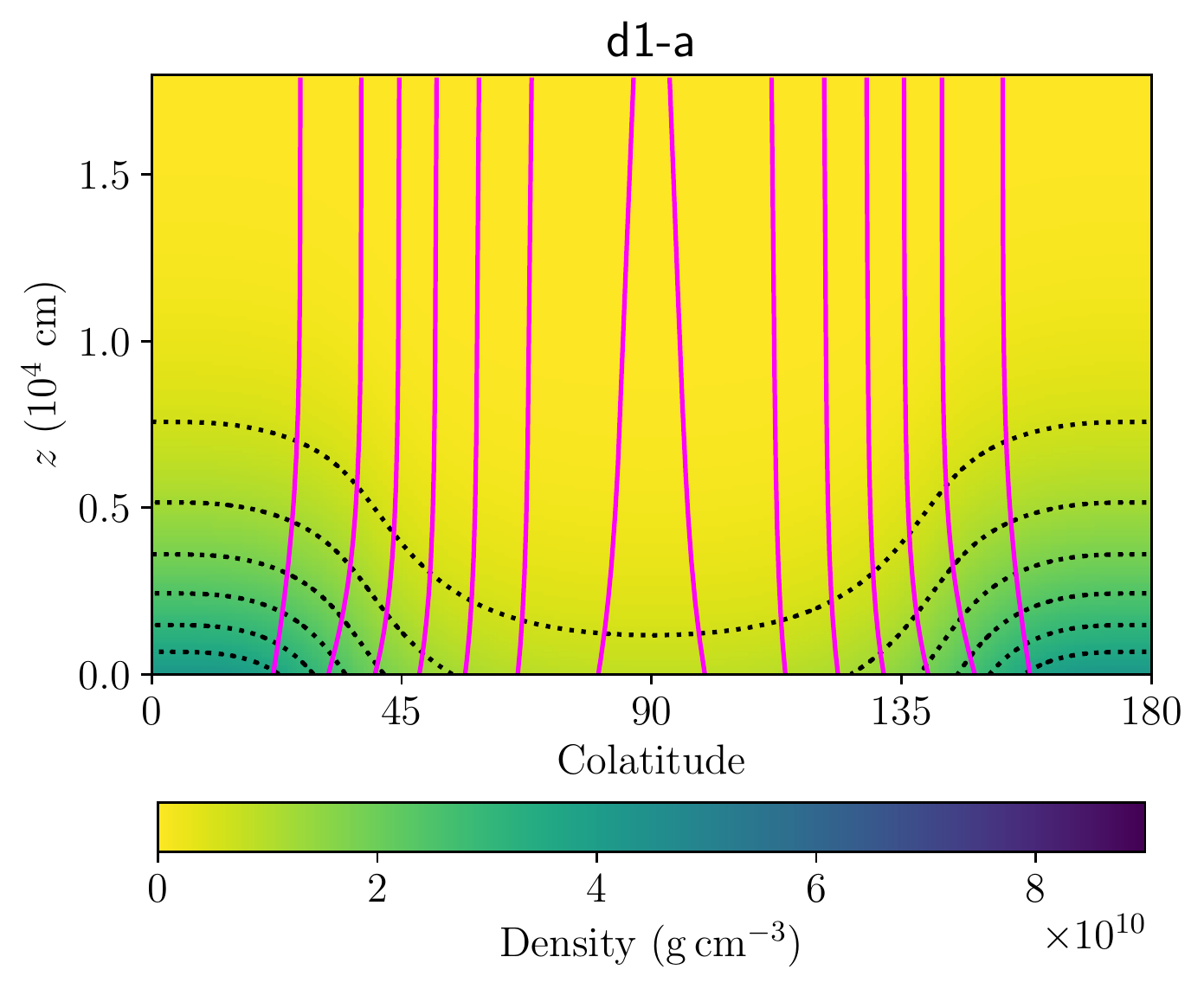}
   \includegraphics[width=0.49\textwidth]{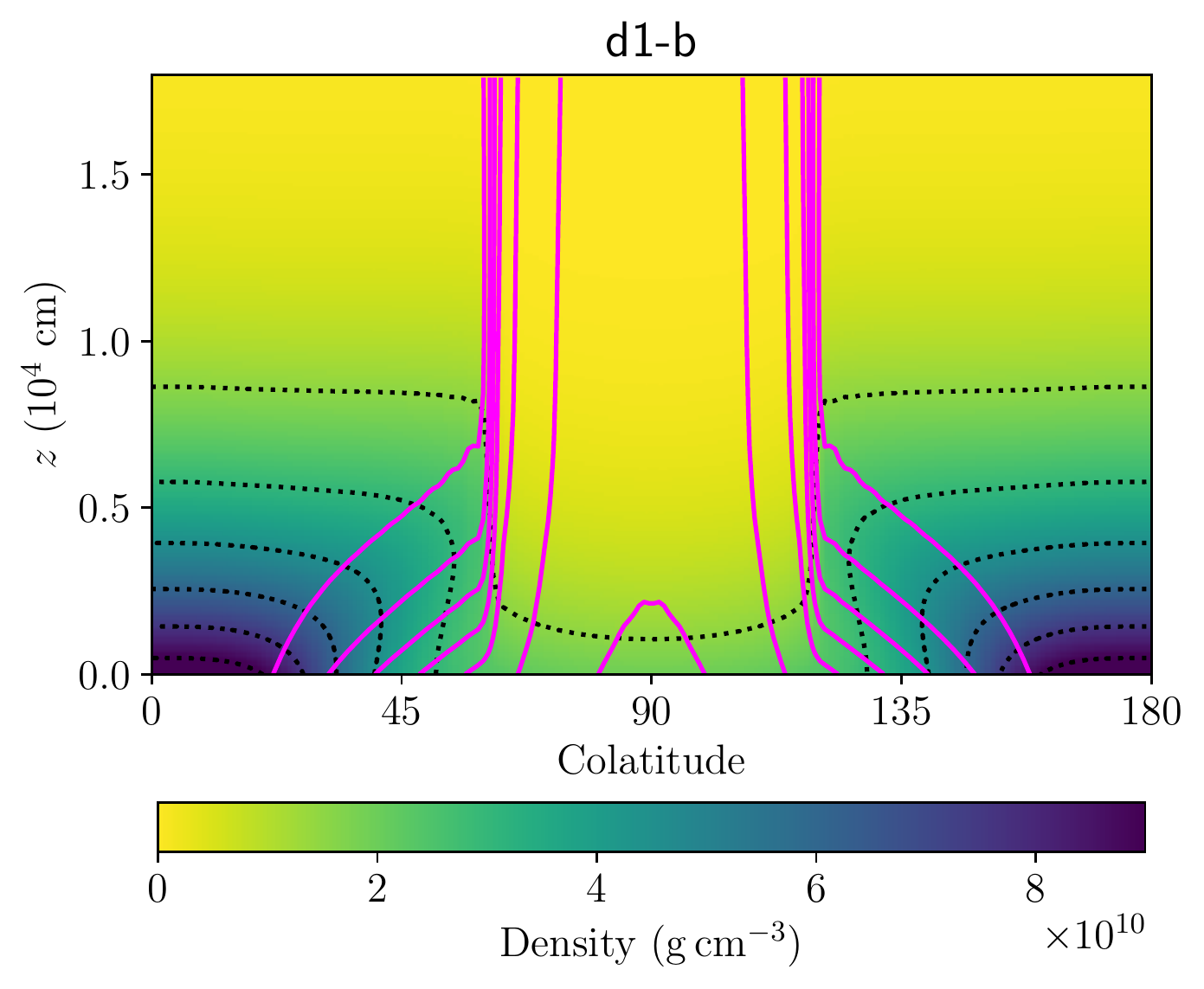}

  \caption{Top: Magnetic field lines of solutions d1-a (left) and d1-b (right) at a distance from the stellar surface. Bottom: Density structures and magnetic field lines of solution d1-a (left) and d1-b (right). The dotted curve indicates the density contour, and the solid line is the magnetic field line.}
    \label{fig:d1Psi}
\end{figure*}

\begin{figure*}
    \centering
   \includegraphics[width=0.49\textwidth]{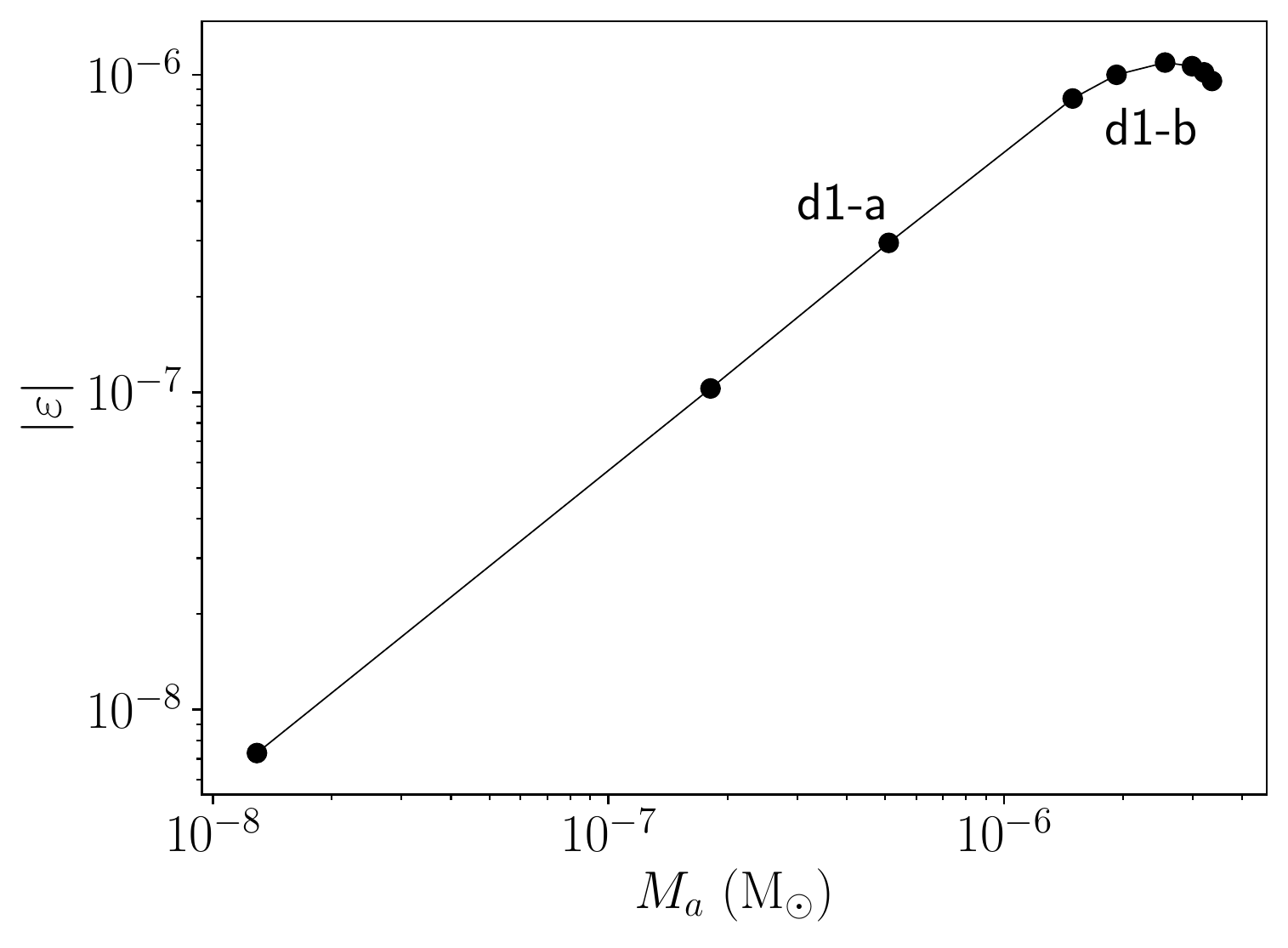}
      \includegraphics[width=0.49\textwidth]{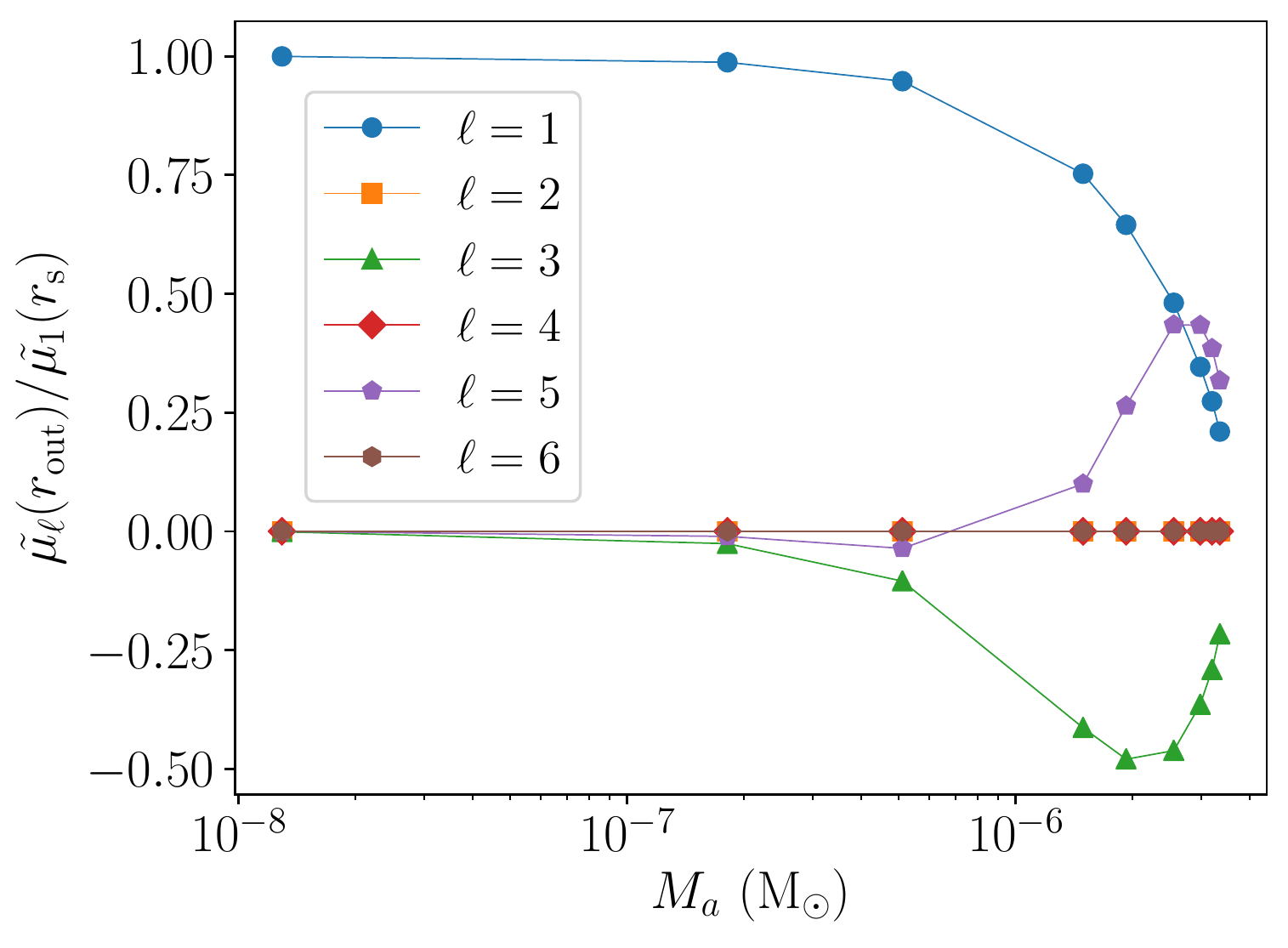}

  \caption{Left: Ellipticity as a function of accreted mass $M_a$. Right: Multipole magnetic moments as a function of accreted mass $M_a$. Magnetic multipole moments at the outer boundary $\tilde{\mu}_\ell(r_{\mathrm{out}})$ are normalized by stellar dipole magnetic moment $\tilde{\mu}_1(r_s)$.}
    \label{fig:d1eps}
\end{figure*}

First, we show the numerical results for the case where a neutron star has a dipole field (model d1 in Table \ref{tab:param}). The matter accretes on to the magnetic poles at the north and south poles, and this configuration is the same as in previous studies (\citealt{Payne_Melatos_2004}, \citealt{Priymak_et_al_2011},  \citealt{Suvrov_Melatos_2019}, \citealt{Suvorov_Melatos_2020}). Figure~\ref{fig:d1Psi} displays the magnetic field lines (solid curves) and density contours (dotted curves) of solutions d1-a (left) and d1-b (right). Figure~\ref{fig:d1eps} displays the ellipticity (left) and the normalized magnetic moments (right) as a function of the accreted mass $M_a$. In the left panel of Fig.~\ref{fig:d1eps}, solutions d1-a and d1-b are shown. The accreted masses of solutions d1-a and d1-b are $M_a \sim 5.1\times 10^{-7}\mathrm{M_\odot}$ and $M_a \sim 1.9\times10^{-6}\mathrm{M_\odot}$, respectively. Solution d1-a is typical of a magnetic mountain with a dipole field, and solution d1-b has the almost maximum ellipticity in the solution sequence. The ellipticity is always negative $(\varepsilon < 0)$ because the matter is accreted on to the magnetic pole (both north and south poles), and the mountains are formed on both poles. Since we fix the magnetic field lines $\Psi$ at the stellar surface, the dipole field is deformed because of the accretion, and the matter is compressed (bottom panels in Fig.~\ref{fig:d1Psi}). As seen in the left panel of Fig.~\ref{fig:d1eps}, the absolute value of the ellipticity increases monotonically up to a certain solution and then decreases. This behaviour is the same as in previous studies (see fig.~4 in \citealp{Priymak_et_al_2011}). The stellar dipole magnetic field sustains the almost maximum mountain in solution d1-b. Beyond the solution, however, the mountain is flattened due to the accreted matter, and the ellipticity decreases. 

The right panel in Fig.~\ref{fig:d1eps} shows the normalized magnetic multipole moments at the outer boundary. The magnetic multipole moments at the outer boundary $\tilde{\mu}_\ell(r_{\mathrm{out}})$ are normalized by the stellar dipole magnetic moment at the surface $\tilde{\mu}_1(r_s)$.  As the accreted mass increases, the dipole magnetic field is buried, and the dipole magnetic moment $(\ell = 1)$ decreases. Approximately 30\% of the dipole moment decreases as a result of accretion when the ellipticity reaches the maximum at solution d1-b. The distorted magnetic field is no longer dipole. As the dipole field is distorted, the dipole moment is transformed into multipole moments (see \citealt{Priymak_et_al_2011}). Since the neutron star has only a dipole ($\ell = 1$) magnetic moment and accretion is equatorially symmetric in this model, there are no even magnetic moments ($\ell = 2, 4, 6$) in Fig.~\ref{fig:d1eps}.

\begin{figure*}
    \includegraphics[width=0.47\textwidth]{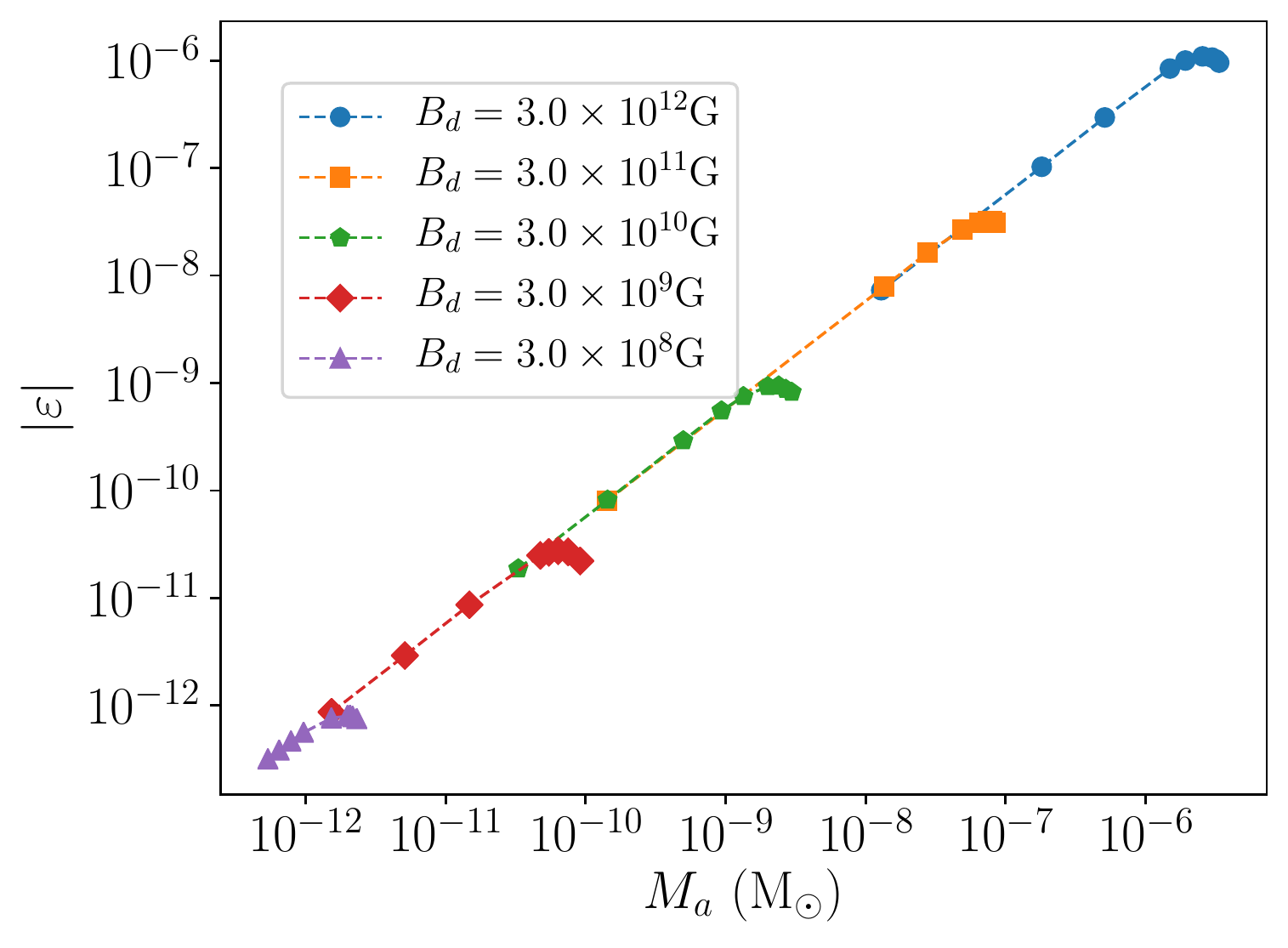}
    \includegraphics[width=0.47\textwidth]{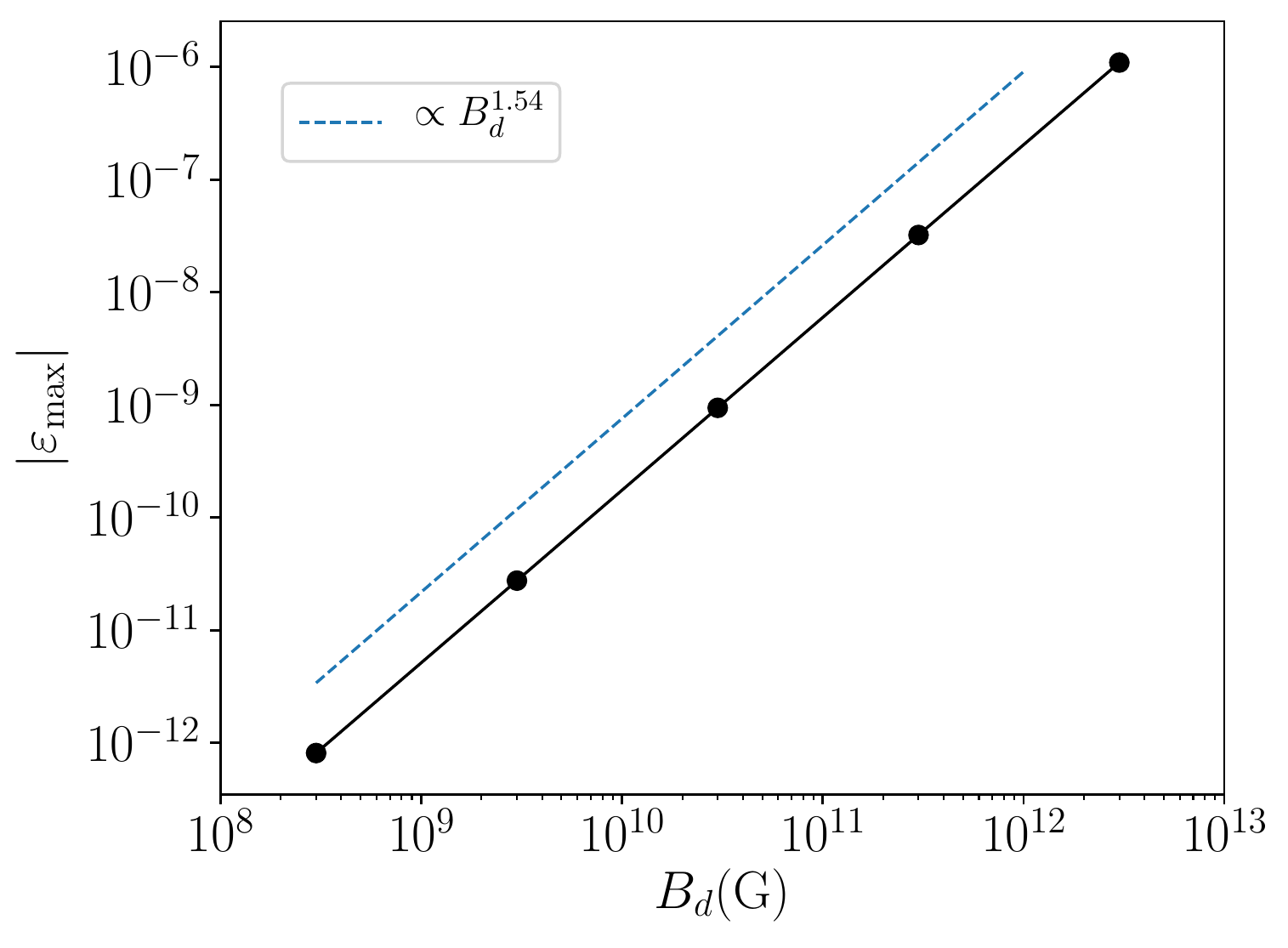}
    \caption{Left: Ellipticity as a function of accreted mass $M_a$. Right: Maximum ellipticity $|\epsilon_{\max}|$ as a function of $B_d$.}
    \label{fig:ellipticities}
\end{figure*}

 The numerical solutions are approximately dimensionless because we use a polytropic equation of state and assume a solid surface. There is expected to be a scaling relation between the magnetic field and the ellipticity. We plot the values of ellipticities in the left panel of Fig.~\ref{fig:ellipticities} by changing the strength of $B_d$ from $3 \times 10^{8}$G to $3 \times 10^{12}$G. We also plot the absolute value of the maximum ellipticity $|\epsilon_{\max}|$ as a function of $B_d$ in the right panel of Fig.~\ref{fig:ellipticities}. We find that the value of the maximum ellipticity scales as follows:
\begin{align}
    |\epsilon_d
    | \sim \left( \frac{B_d}{B_0} \right)^{1.54}.
    \label{eq:eps}
\end{align}
This scaling relation is different from that of a magnetized deformed star whose ellipticity scales as $|\epsilon| \sim B^2$ (\citealt{Haskell_et_al_2008}; \citealt{Mastrano_et_al_2011}; \citealt{Mastrano_et_al_2015}). This difference might be due to the compression of density in the case of magnetic mountains. In the previous papers, the magnetized deformed star is calculated perturbatively. However, the magnetically confined mountain in this paper is compressed self-consistently, and the compression might change the scaling relation. Therefore, the maximum ellipticity of the dipole field model $\epsilon_{d}$ is scaled as 
\begin{align}
    |\epsilon_{d}| \sim 1\times 10^{-6}\left( \frac{B_d}{3.0\times 10^{12}~\mathrm{G}} \right)^{1.54}.
    \label{eq:eps_d}
\end{align}
If we consider a typical millisecond pulsar with $B_d/B_0 = 10^{-4}$, the ellipticity becomes six orders of magnitude smaller (see the right panel in Fig.~\ref{fig:ellipticities}). 

\subsection{Octupole magnetic field burial}

\begin{figure*}
    \centering

   \includegraphics[width=0.47\textwidth]{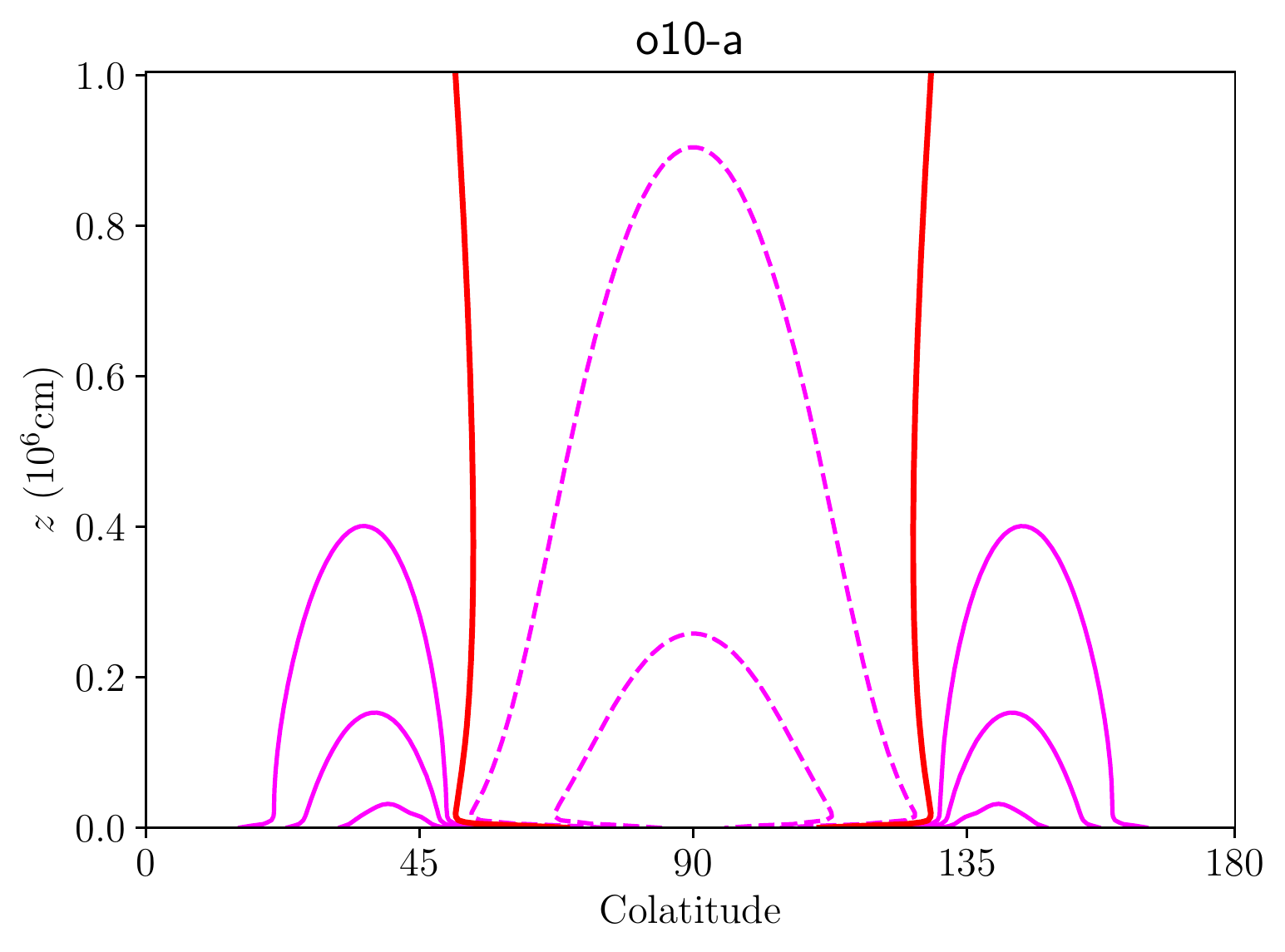}
   \includegraphics[width=0.47\textwidth]{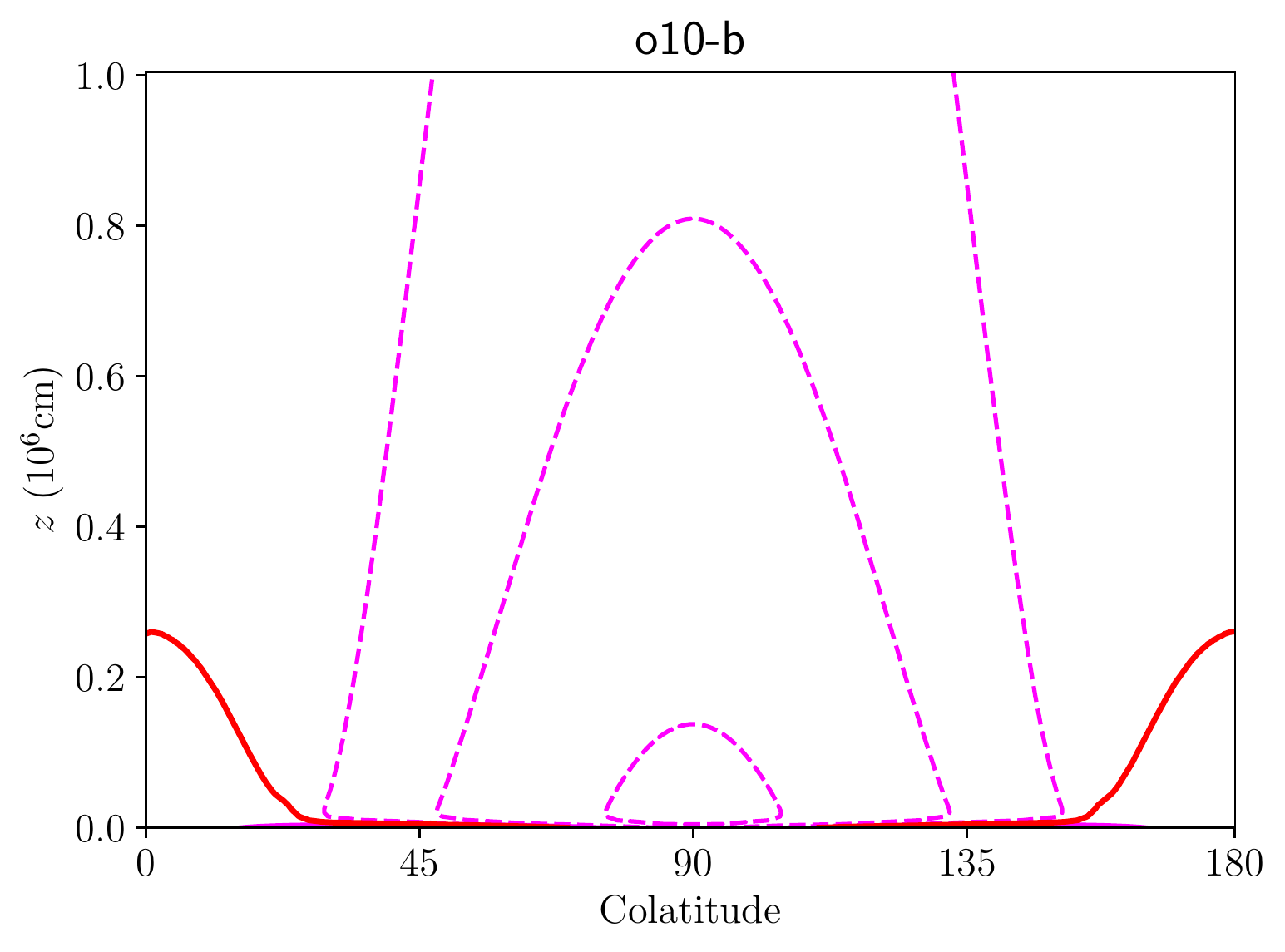}

   \includegraphics[width=0.47\textwidth]{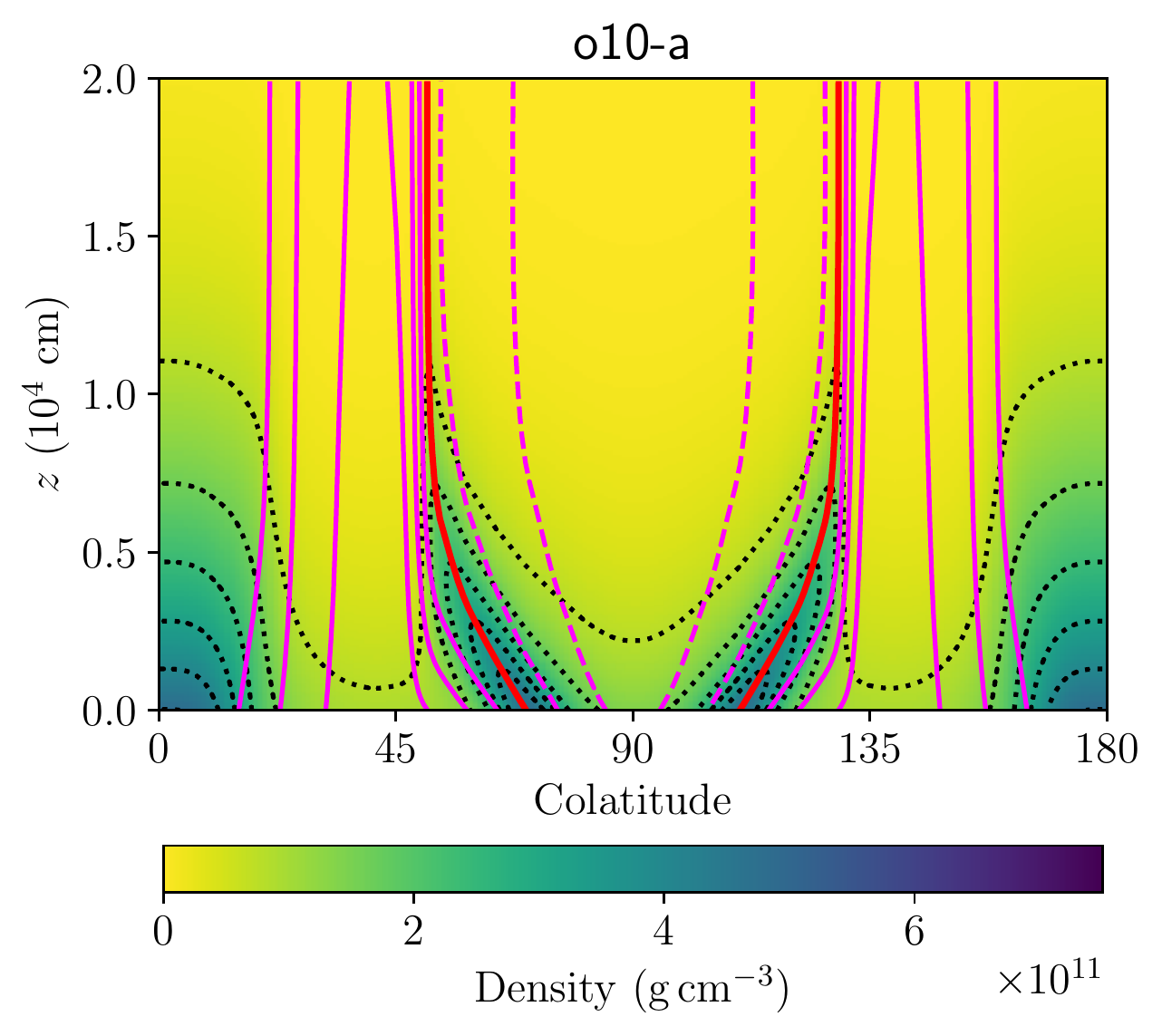}
   \includegraphics[width=0.47\textwidth]{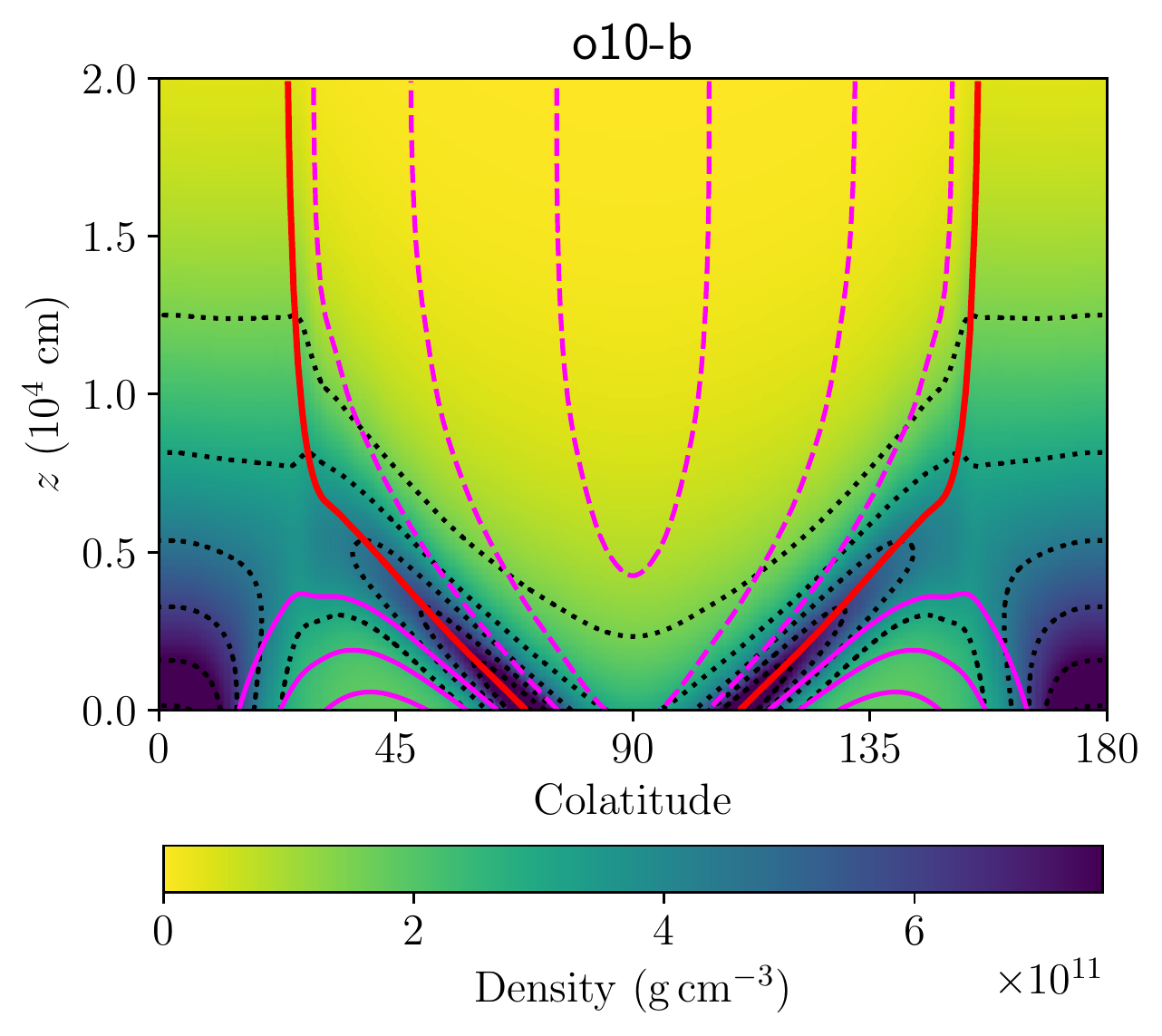}

  \caption{Top: Magnetic field lines of solutions o10-a (left) and o10-b (right) at a distance from the stellar surface. The solid line is the positive magnetic field line, and the dashed line is the negative magnetic field line. The thick red line denotes the zero magnetic flux ($\Psi = 0$), and the direction of the magnetic field line is reversed at this line. Bottom: Density structure and magnetic field lines of solution o10-a (left) and o10-b (right). The dotted curve indicates the density contour, the solid line is the positive magnetic field line, and the dashed line is the negative magnetic field line. The thick red line denotes the zero magnetic flux ($\Psi = 0$).}
    \label{fig:o10-Psi}
\end{figure*}

\begin{figure*}
    \centering
   \includegraphics[width=0.47\textwidth]{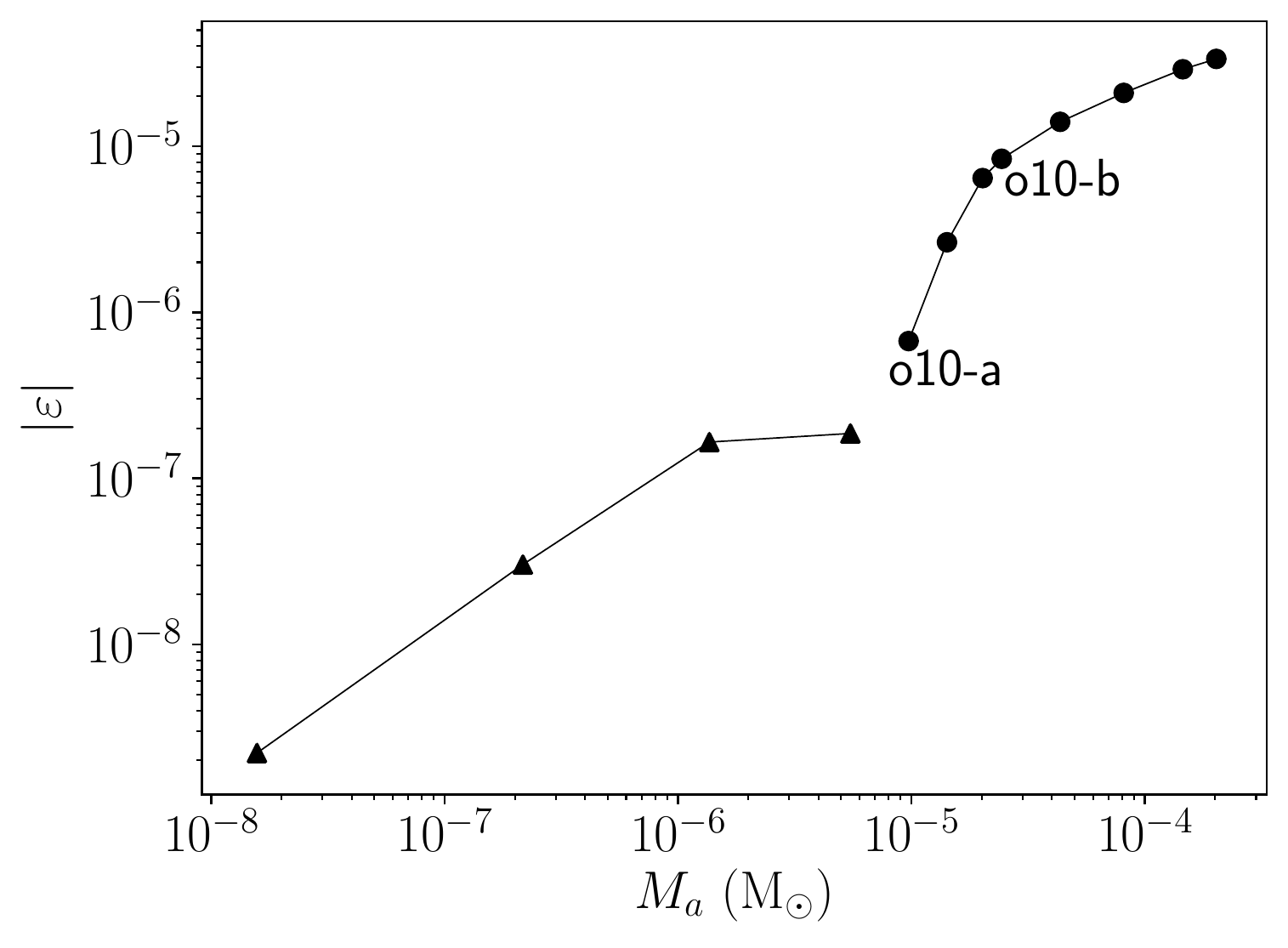}
      \includegraphics[width=0.47\textwidth]{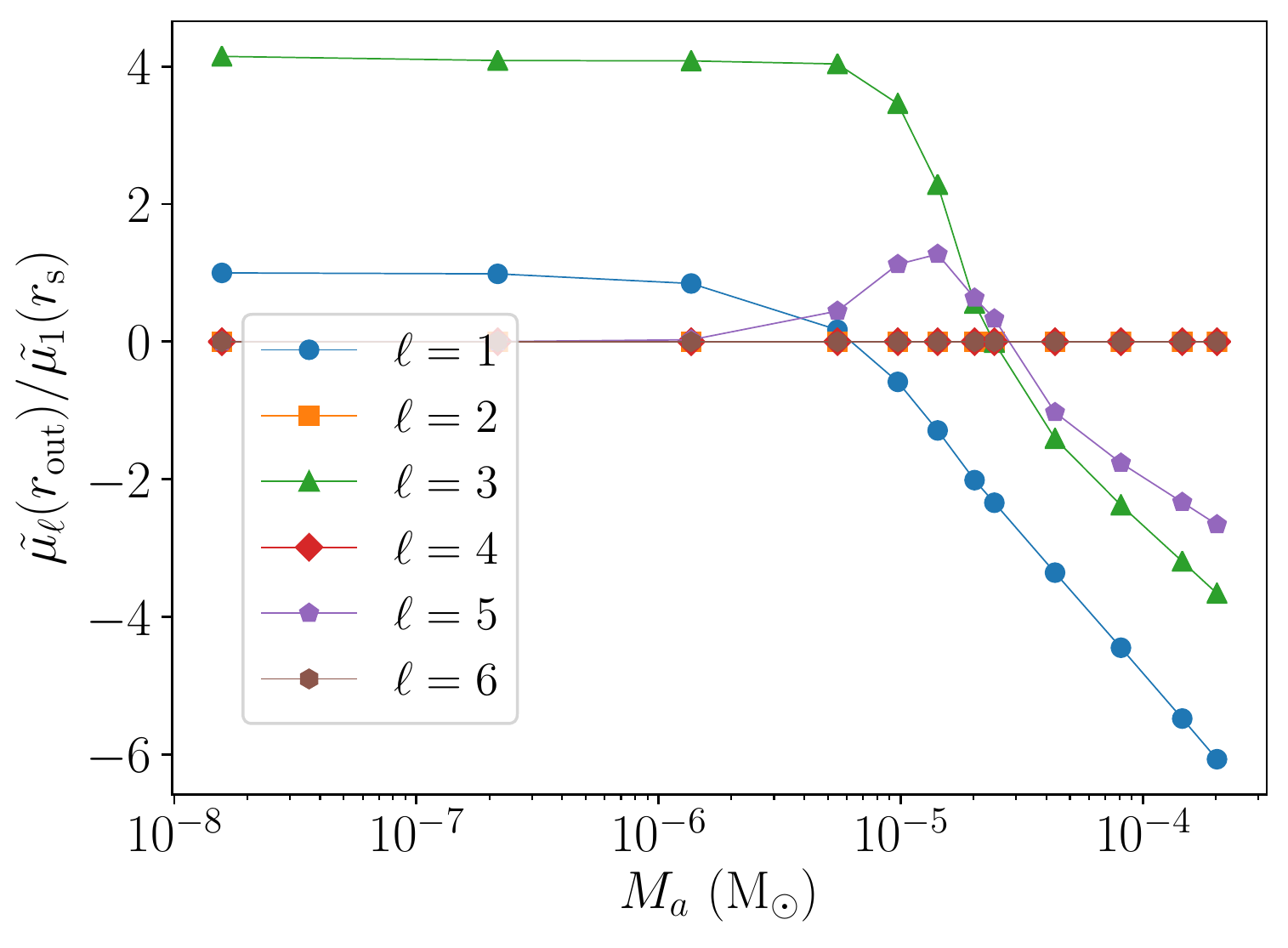}

  \caption{Left: Ellipticity as a function of accreted mass $M_a$. The triangle denotes $\varepsilon > 0$, while the circle denotes $\varepsilon < 0$. Right: Normalized magnetic moments function as accreted mass $M_a$. Magnetic multipole moment at the outer boundary $\tilde{\mu}_\ell (r_{\mathrm{out}})$ is normalized by stellar dipole magnetic moment $\tilde{\mu}_1 (r_s)$. }
    \label{fig:o10-eps}
\end{figure*}

Next, we show the results for the case where the neutron star has dipole and strong octupole magnetic fields (model o10 in Table~\ref{tab:param}). The strength of the octupole magnetic field is ten times larger than that of the dipole magnetic field at the surface. Figure \ref{fig:o10-Psi} displays the structure of solution o10-a (left) and solution o10-b (right). Figure \ref{fig:o10-eps} shows the ellipticity (left) and the magnetic multipole moments (right) as a function of the total accreted mass. In the left panel of Figure \ref{fig:o10-eps}, solutions o10-a and o10-b are shown. The accreted masses of solutions o10-a and o10-b are $M_a \sim 9.7 \times 10^{-6}\mathrm{M_\odot}$ and $M_a \sim 2.4 \times10^{-5}\mathrm{M_\odot}$, respectively. In the case of multipole magnetic fields, there are more than two magnetic poles, and in the case of the octupole magnetic fields, there are four magnetic poles, as seen in Fig.~\ref{fig:o10-Psi}. Due to the functional form for $F(\Psi)$, the matter accretes on to all magnetic poles of the octupole magnetic field, and four mountains are formed in this model. Since the mass accretes on to these four magnetic poles, the signs of the ellipticity of some solutions are positive. The system is equatorially symmetric because the neutron star has only odd ($\ell = 1$ and $\ell = 3$) magnetic moments, as seen in Fig~\ref{fig:o10-eps}. 

Figure \ref{fig:o10-eps} shows that the absolute value of the ellipticity increases monotonically. This behaviour of ellipticity is different from the dipole magnetic field case in Fig.~\ref{fig:d1eps}. The maximum value of ellipticity for $B_d = 3.0\times 10^{12}~\mathrm{G}$ and the strong octupole field $\varepsilon_o$ case scales as follows:
\begin{align}
|\varepsilon_o| \sim 3 \times 10^{-5} \left( \frac{B_o}{3.0 \times 10^{13}~\mathrm{G}} \right)^{1.54}.
        \label{eq:eps_o}
\end{align}

The right panel of Fig.~\ref{fig:o10-eps} displays the normalized magnetic multipole moments as a function of the accreted mass. As the total accreted mass increases, the value of the octupole moment ($\ell=3$) decreases. The dipole moment ($\ell=1$) also decreases, becoming negative beyond a certain solution. As the value of the octupole moment decreases, the negative dipole moment further decreases. Although the higher-order moments ($\ell =5$) slightly increases,  the negative dipole moment becomes dominant. Finally, the absolute value of the dipole moment is larger than that of the octupole and $\ell = 5$ moments (solution o10-b). This means that the octupole magnetic field is buried completely, and the neutron star appears to have a (negative) dipole magnetic field outside the mountain (right panel in Fig~.\ref{fig:o10-Psi}). The octupole magnetic field is transformed into the negative dipole magnetic field. 

\subsection{Quadrupole magnetic field burial}

\begin{figure*}
    \centering

   \includegraphics[width=0.47\textwidth]{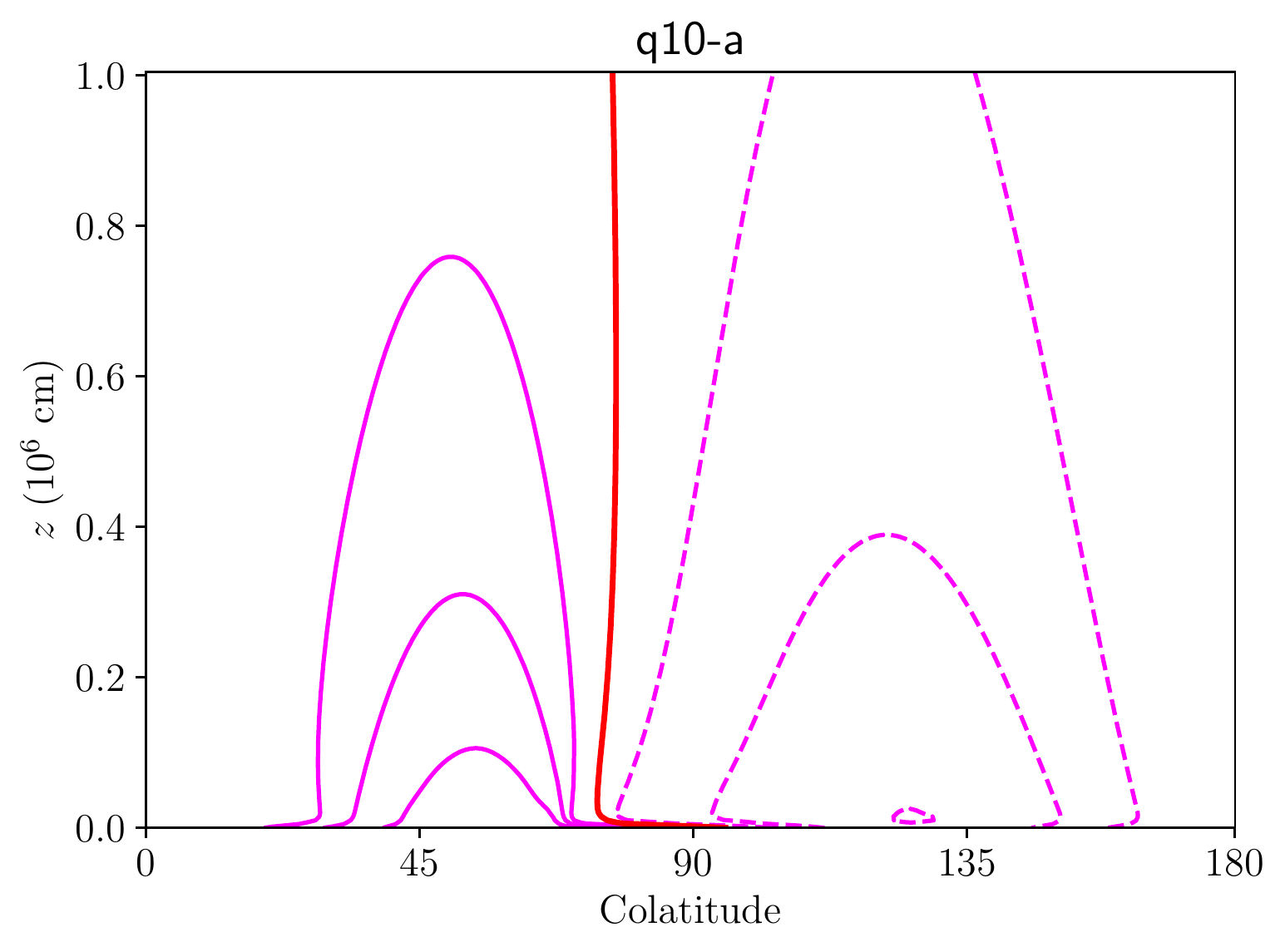}
   \includegraphics[width=0.47\textwidth]{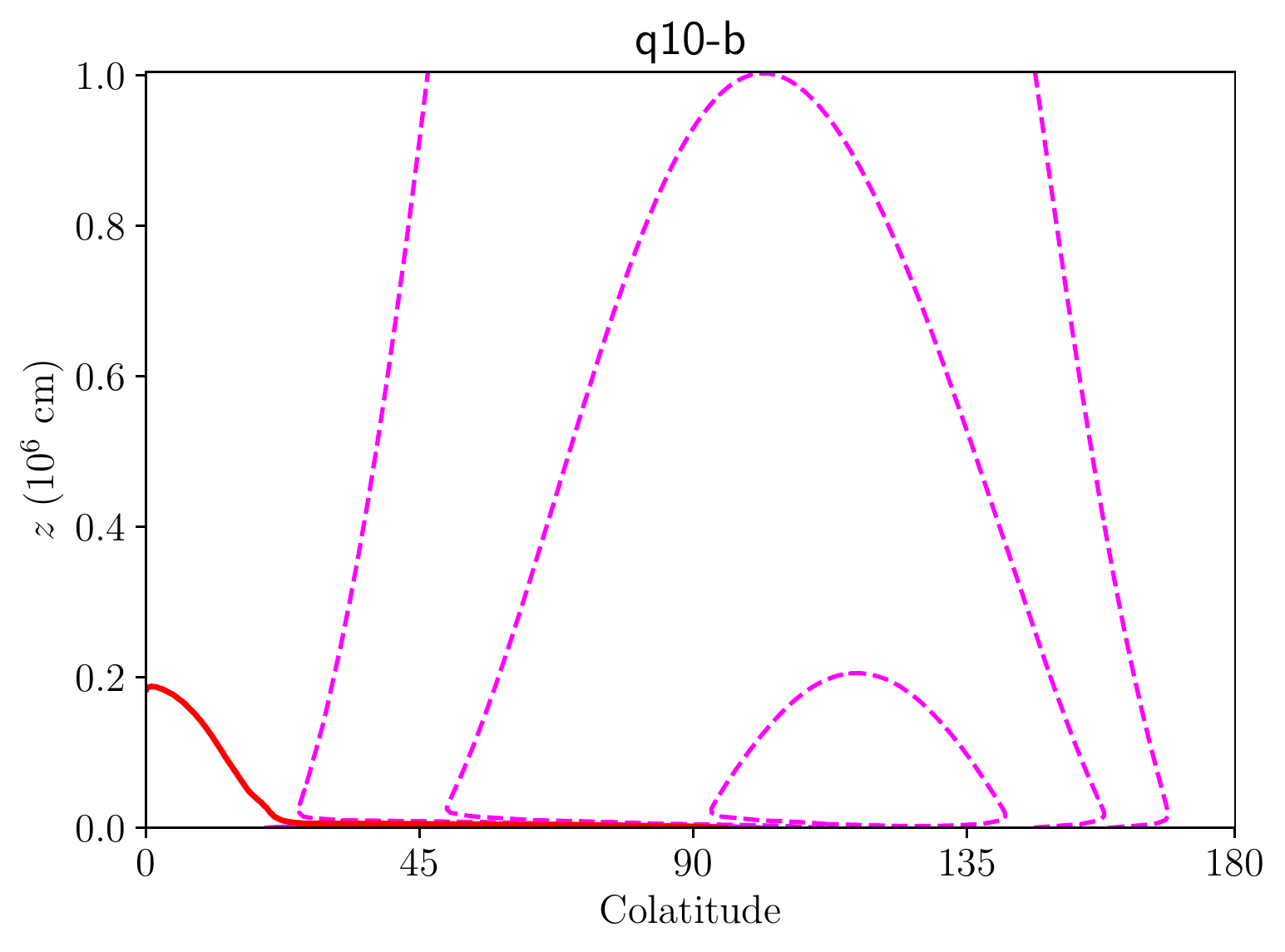}

   \includegraphics[width=0.47\textwidth]{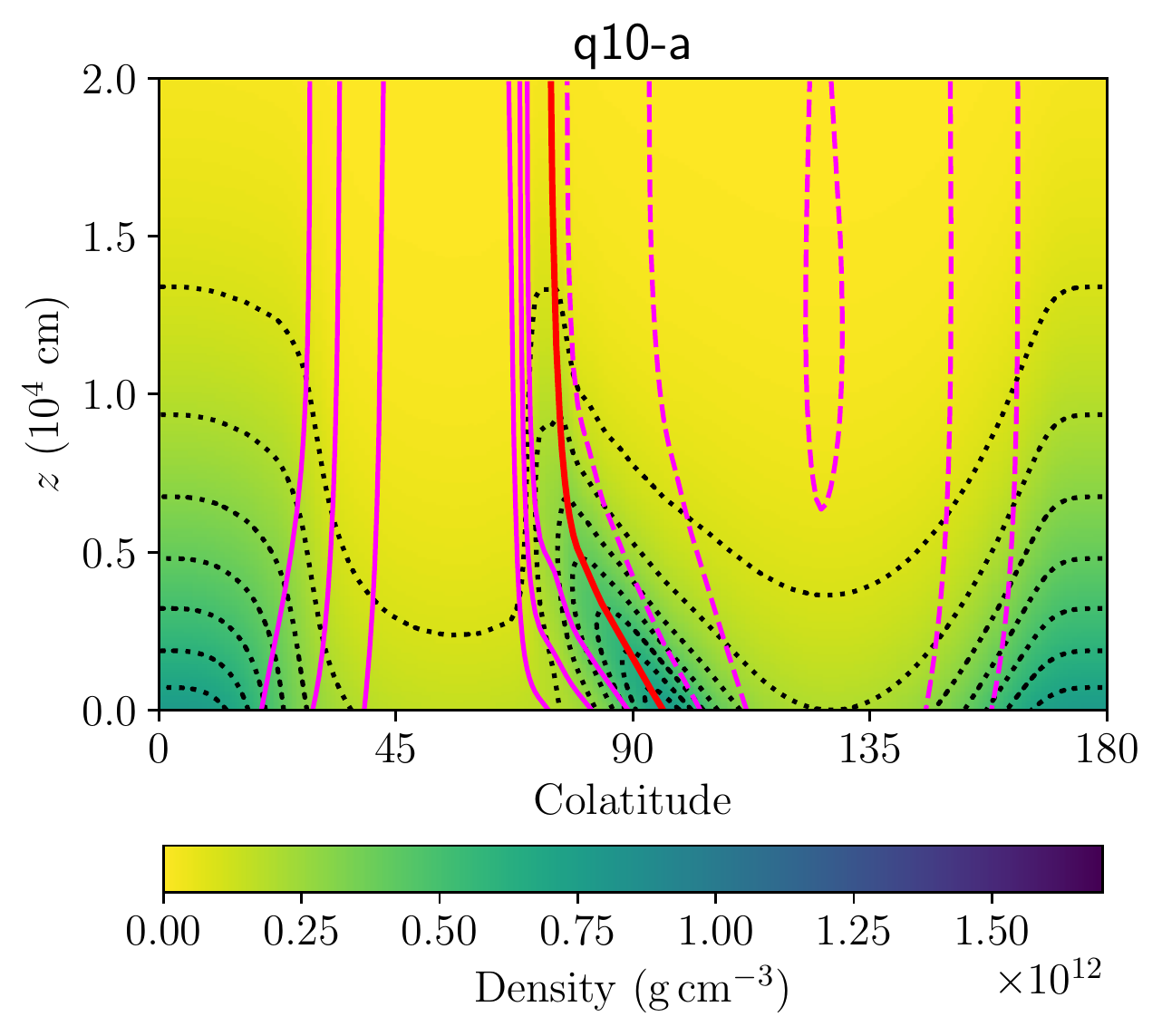}
   \includegraphics[width=0.47\textwidth]{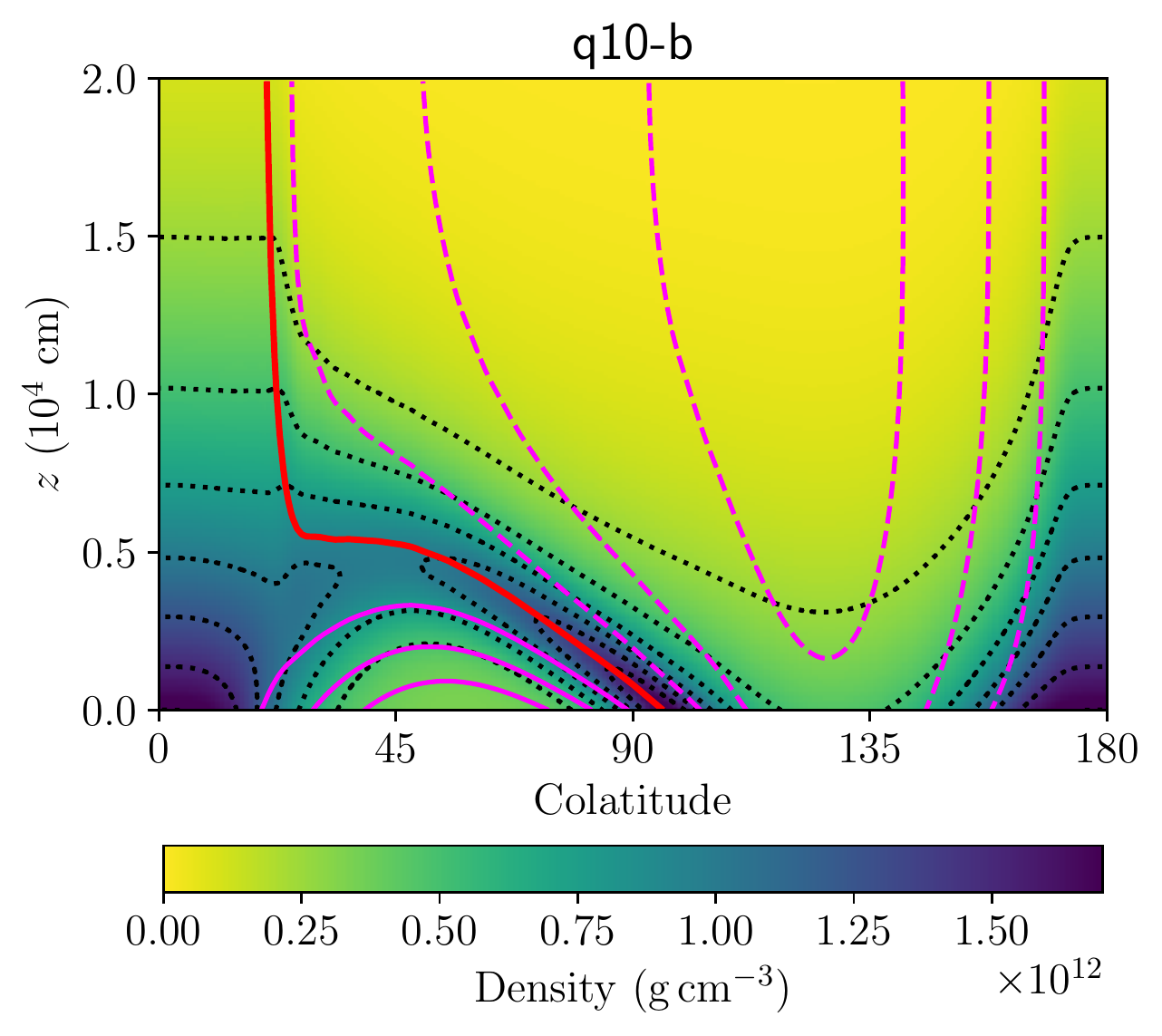}

  \caption{As per Fig.~\ref{fig:o10-Psi} but model q10.}
    \label{fig:q10-psi}
\end{figure*}

\begin{figure*}
    \centering
    \includegraphics[width=0.47\textwidth]{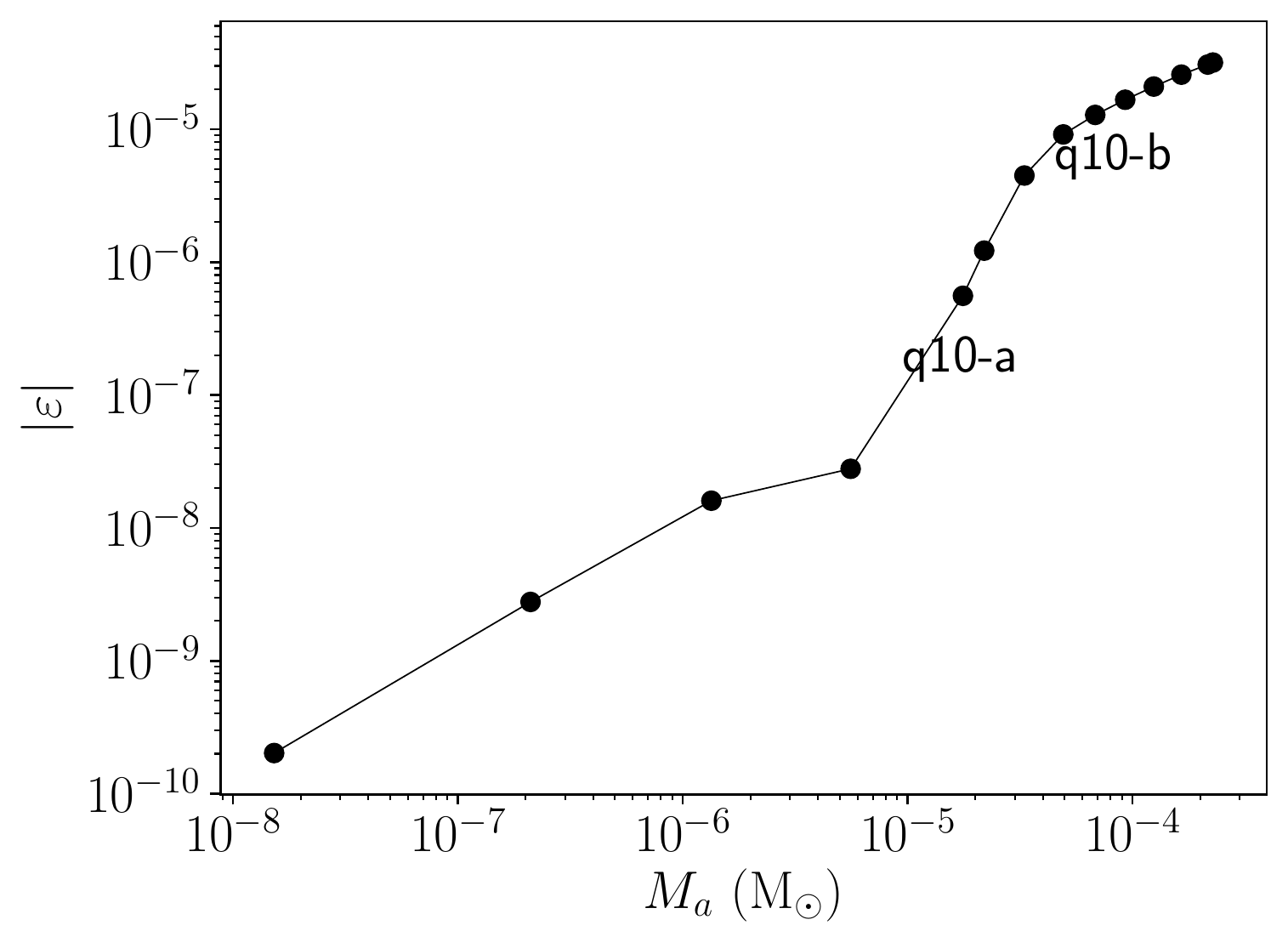}
    \includegraphics[width=0.47\textwidth]{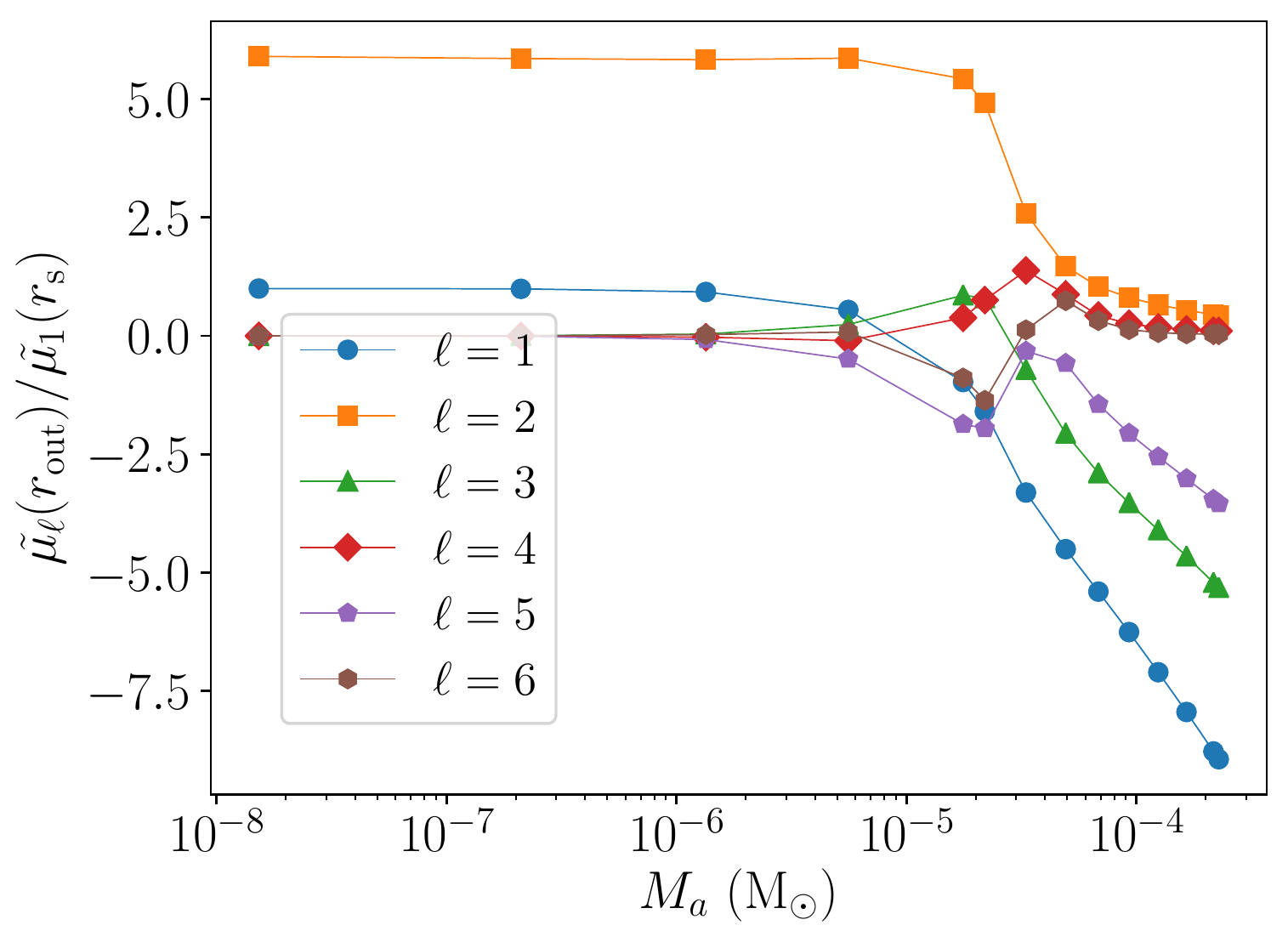}

  \caption{As per Fig.~\ref{fig:o10-eps} but model q10.}
    \label{fig:q10-eps}
\end{figure*}

We show the results for the case where the neutron star has dipole and strong quadrupole magnetic fields (model q10 in Table~\ref{tab:param}). Figure \ref{fig:q10-psi} displays the magnetic field lines of the solutions, and Figure \ref{fig:q10-eps} shows the ellipticity (left) and normalized magnetic multipole moments (right) as functions of the total accreted mass. In the left panel of Figure \ref{fig:q10-eps}, solutions q10-a and q10-b are shown. The accreted masses of solutions q10-a and q10-b are $M_a \sim 1.8\times 10^{-5}\mathrm{M_\odot}$ and $M_a \sim 4.9\times10^{-5}\mathrm{M_\odot}$, respectively. As seen in the bottom panels of Fig.~\ref{fig:q10-psi}, there are three magnetic poles and three mountains on the surface. The equatorial symmetry breaks because the quadrupole magnetic field is not equatorially symmetric. The northern hemisphere has two magnetic poles, while the southern hemisphere has only one magnetic pole.

Figure~\ref{fig:q10-eps} shows that the ellipticity always increases monotonically, and we do not find the turnover behaviour seen in the dipole models. Using the scaling relation in equation \eqref{eq:eps}, the maximum ellipticity for $B_d = 3.0\times 10^{12}~\mathrm{G}$ with strong octupole field $\varepsilon_q$ scales approximately as below:
\begin{align}
|\varepsilon_q| \sim 3 \times 10^{-5} \left( \frac{B_q}{3.0 \times 10^{13}~\mathrm{G}} \right)^{1.54}.
    \label{eq:eps_q}
\end{align}

As the accreted mass increases, the quadrupole magnetic field is buried, and the quadrupole moment ($\ell = 2$) decreases, as seen in the right panel of Fig.~\ref{fig:q10-eps}. The magnetic dipole moment also decreases, and the value becomes negative beyond a certain point. Although the $\ell = 4$ and $\ell =6$ multipole moments increase slightly, the negative dipole moment further decreases. Finally, the negative dipole moment becomes dominant, and one mountain and one magnetic pole are buried completely. It appears that the quadrupole magnetic field is buried and disappears, while the negative dipole magnetic field appears. 

\subsection{Effect of toroidal magnetic fields}

Finally, we calculate solutions with toroidal magnetic fields. Since we use the functional form for $S(\Psi)$ as in equation \eqref{eq:S_Psi}, the toroidal magnetic field is confined within the closed-loop of the mountains. To express the strength of the toroidal magnetic field, we use the maximum value of the ratio of the toroidal magnetic field to the poloidal magnetic field $B_\varphi / B_p$, whose value increases by increasing $S_0$ in equation \eqref{eq:S_Psi}.  We obtain four solution sequences (o10t-a, o10t-b, q10t-a, and q10t-b) by including the toroidal magnetic field in four respective solutions (o10-a, o10-b, q10-a, and q10-b).

\begin{figure}
    \centering
    \includegraphics[scale=0.54]{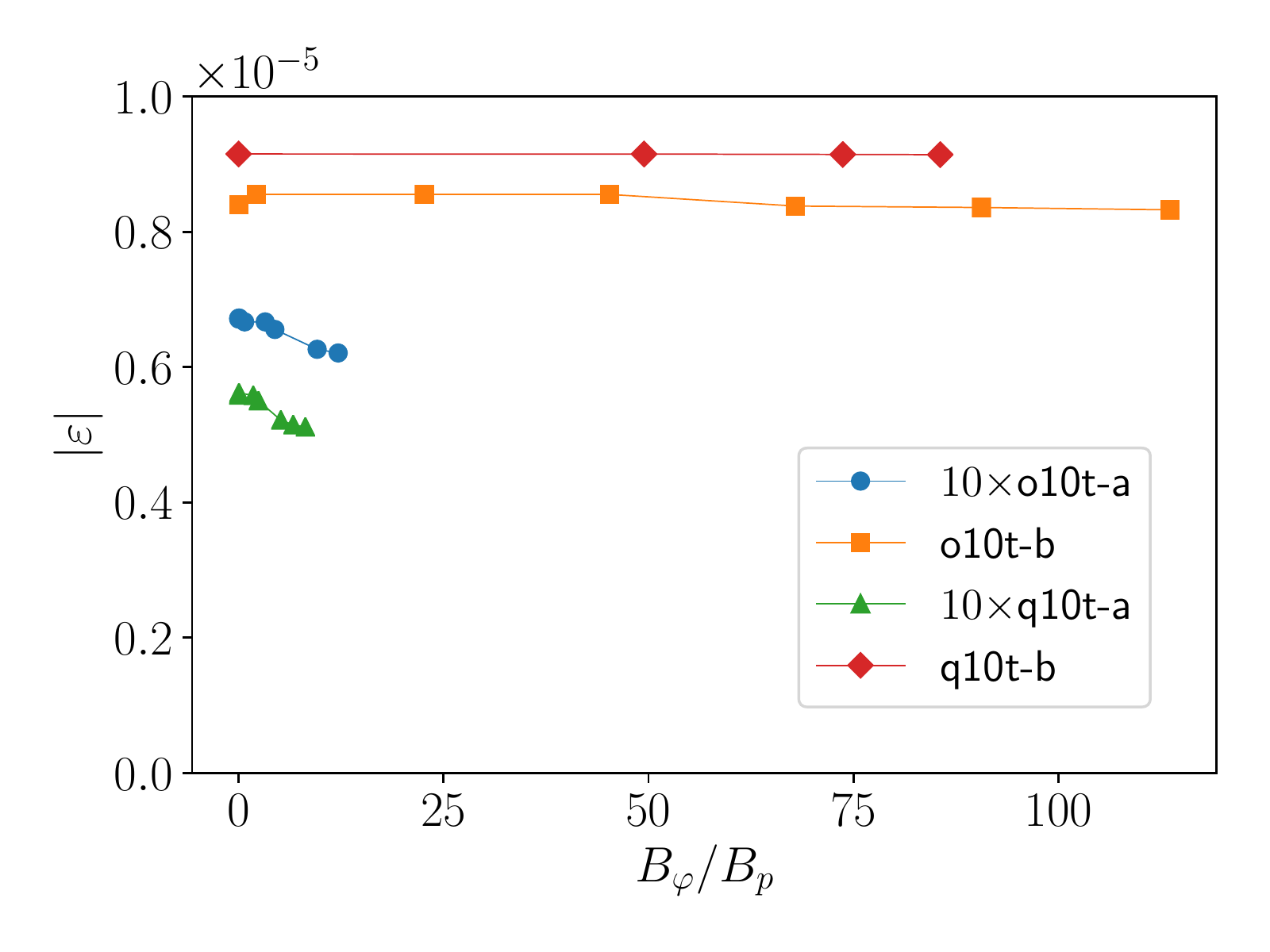}
    \caption{Ellipticity as a function of the ratio of the toroidal magnetic field and the poloidal magnetic field $B_\varphi / B_p$. Each point denotes a solution sequence of o10t-a (blue circles), o10t-b (orange squares), q10t-a (green triangles), and q10t-b (red diamonds), respectively.The values of ellipcities for o10t-a and q10t-a are multiplied by a factor 10.}
    \label{fig:t-eps}
\end{figure}

Figure~\ref{fig:t-eps} displays the effects of toroidal magnetic fields on ellipticity. The ellipticities of four solution sequences (o10t-a, o10t-b, q10t-a, and q10t-b) are shown as a function of the ratio $B_\varphi / B_p$. As the value of $B_\varphi / B_p$ increases, convergence of the solution becomes more difficult. We find a critical solution with the maximum value of $B_\varphi/B_p$, beyond which the iteration does not converge. The solution in Fig.~\ref{fig:t-eps} terminates at the critical solution. The ellipticities of solutions o10t-a and q10t-a decrease by 10\% owing to the toroidal magnetic fields, with the toroidal magnetic fields suppressing the mountains in these models. This is similar to the case of a magnetized star with mixed poloidal and toroidal fields. The deformation due to the poloidal magnetic field is suppressed by the toroidal magnetic field (\citealp{Haskell_et_al_2008, Mastrano_et_al_2011}). By contrast, the ellipticities of solutions o10t-b and q10t-b are almost constant. The multipole magnetic fields in these models are buried, and the toroidal magnetic field does not affect the structures of the mountains. However, the maximum values of $|B_\varphi / B_p|$ in solution sequences o10t-b and q10t-b are much larger than in solution sequences o10t-a and q10t-a. Solution sequences q10t-a and o10t-a terminate at $B_\varphi / B_p \sim 10$, but the ratio reaches $B_\varphi / B_p \sim 100$ in solutions q10t-b and o10t-b. The buried multipole magnetic fields stores intense toroidal magnetic fields. 

\begin{figure*}
    \centering
    \includegraphics[width=0.47\textwidth]{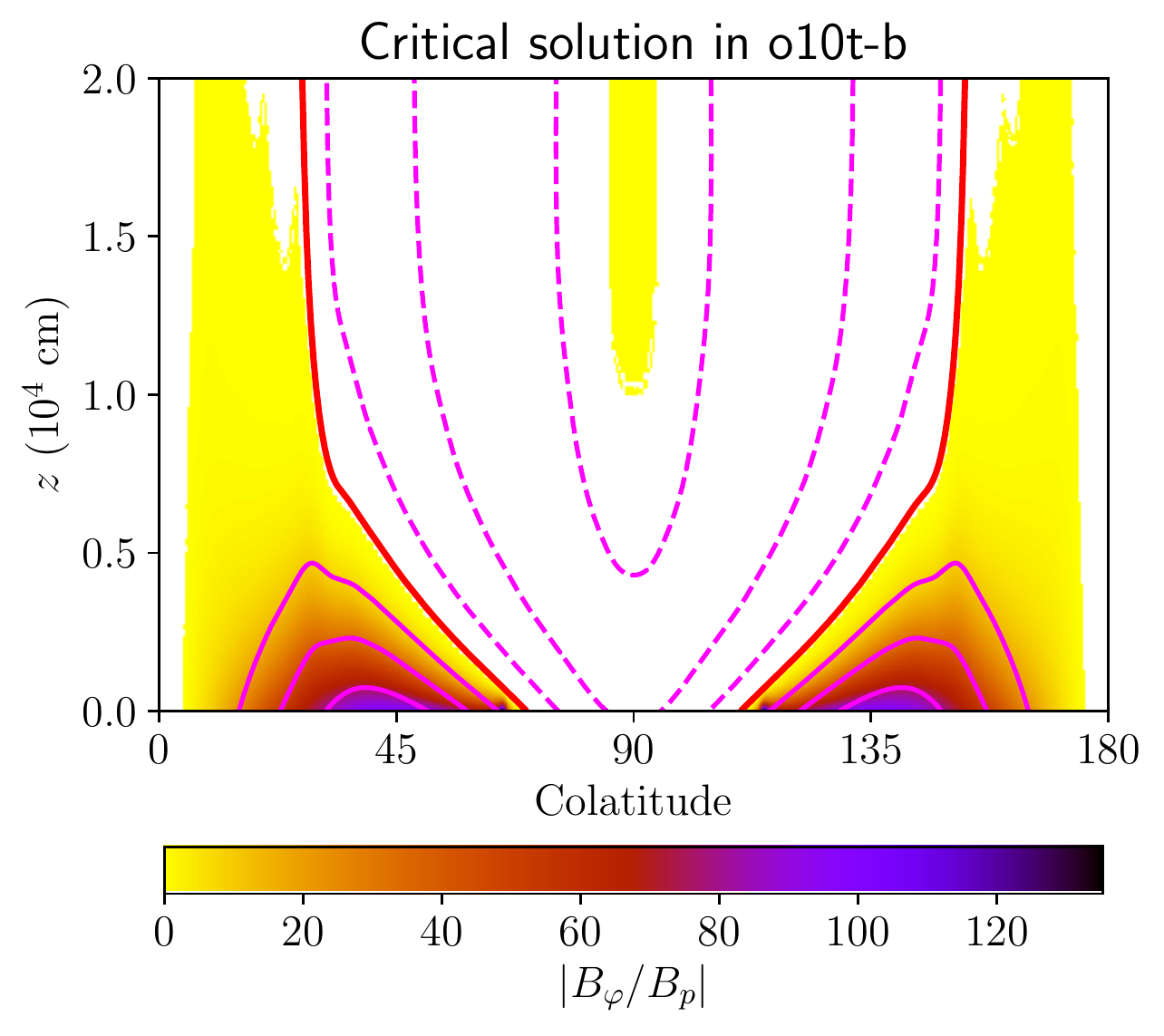}
    \includegraphics[width=0.47\textwidth]{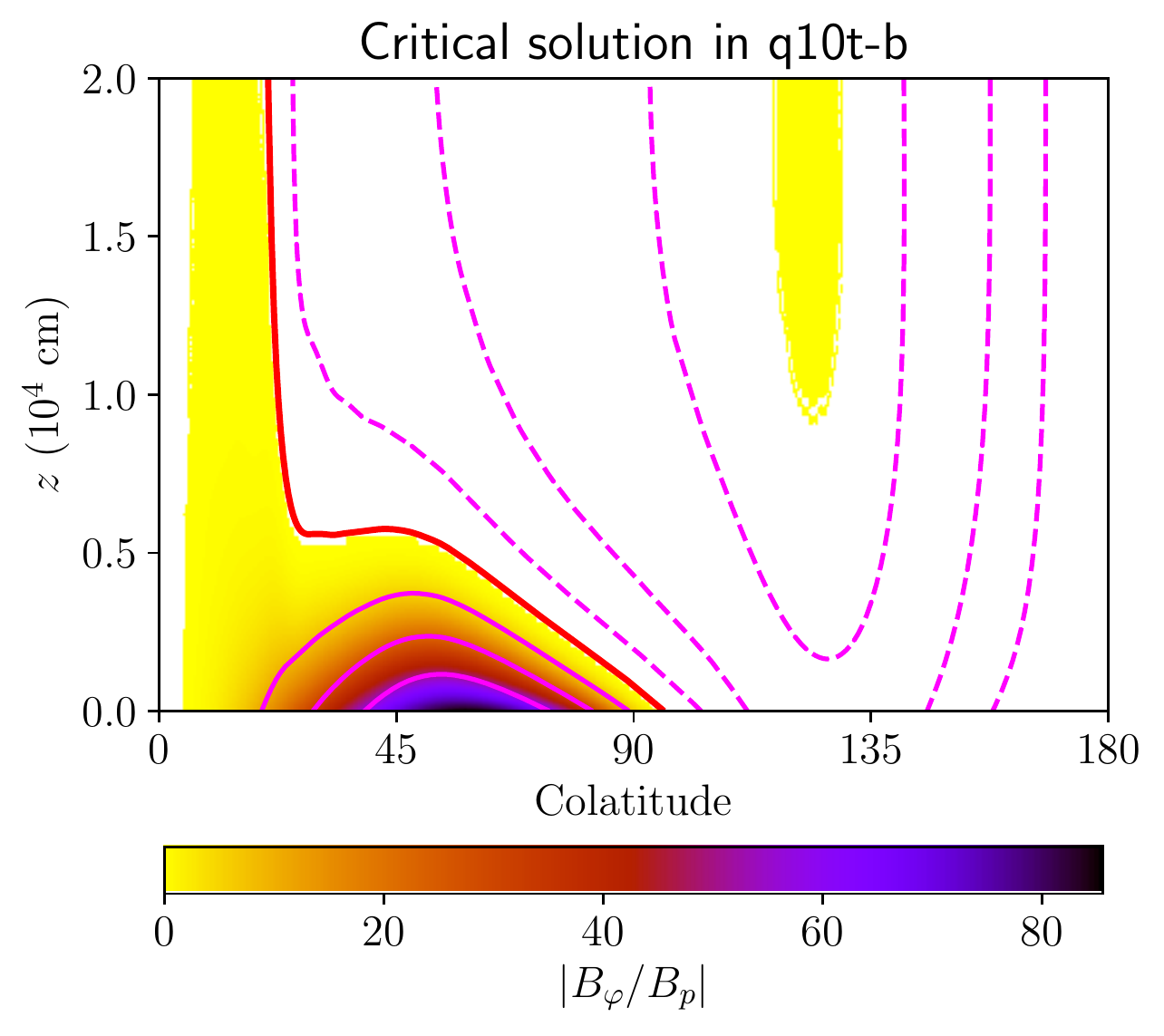}
    \caption{Magnetic field lines and the ratio of the toroidal magnetic field to the poloidal magnetic field $|B_\varphi/B_p|$ (colour map) of critical solutions in solution sequences o10t-b (left) and q10t-b (right). The solid line is the positive magnetic field line, and the dashed line is the negative magnetic field line. The thick red line denotes the zero magnetic flux ($\Psi = 0$).}
    \label{fig:Bphi}
\end{figure*}

Figure~\ref{fig:Bphi} displays the magnetic field lines and the ratio of the toroidal magnetic field to the poloidal magnetic field $|B_\varphi / B_p|$ (colour maps) of critical solutions with the maximum value of $B_\varphi/B_p$. The values of $|B_\varphi / B_p|$ are of the order of 100 inside the closed-loop of the multipole magnetic fields. The multipole magnetic fields buried by the accretion store the intense toroidal magnetic field near the stellar surface. 

\section{Discussion}

\subsection{Stability of the magnetically confined mountain}

The stability of the magnetically confined mountain is important. According to a stability analysis of magnetized stars, a purely poloidal magnetic field or purely toroidal magnetic field is unstable because of the various unstable modes (\citealp{Markey_Tayler_1973, Tayler_1973}). A stable magnetic field configuration should have both poloidal and toroidal components (\citealp{Tayler_1980, Braithwaite_Spruit_2004, Akgun_et_al_2013}). 

If the magnetic field is stable, the magnetically confined mountains would have poloidal and toroidal magnetic fields. \citet{Vigelius_Melatos_2008} performed dynamical simulations of a magnetically confined mountain in three dimensions using the axisymmetric equilibrium solutions of \citet{Payne_Melatos_2004} as initial conditions. They found that the axisymmetric solutions are susceptible to non-axisymmetric instability, and the toroidal field does not qualitatively alter the system's stability. Magnetically confined mountains are generally unstable as shown by numerical dynamical calculations (\citealp{Payne_Melatos_2007, Vigelius_Melatos_2009, Mukherjee_et_al_2013a, Mukherjee_et_al_2013b}). However, \cite{Vigelius_Melatos_2008} also found that the hydrodynamic structure is reconfigured globally and eventually reaches a new, non-axisymmetric mountain confined to the magnetic poles. Mountains with strong multipole magnetic fields may also be unstable, owing to the non-axisymmetric modes, but they will be reconstructed and become new non-axisymmetric mountains.

If the mass accretion rate is high and the timescale of the fallback is faster than the magnetohydrodynamic (MHD) instability timescale, the mountain would be sustained by the mass accretion. The MHD instability timescale of the mountain $t_A$ is estimated as $t_A \sim (r \sqrt{4\pi \rho} / B)$. If the specifications of the mountain $r \sim 10^4 \mathrm{cm}$, $\rho \sim 10^{11}\mathrm{g~cm^{-3}}$, and $B = 10^{12}\mathrm{G}$ are taken as typical values of a magnetic mountain, then the timescale would be of the order of $10^{-2}~\mathrm{s}$. On the other hand, the fallback accretion typically sets in at $\sim10$ s after the explosion. The total fallback mass is typically $\sim10^{-4}-10^{-2}~\mathrm{M}_{\odot}$, depending on the progenitor (e.g., \citealp{2012ApJ...757...69U,2016ApJ...821...69E}). Then, a mountain with mass $M_a = 10^{-7} - 10^{-5}~\mathrm{M}_\odot$ could be formed within $10^{-3}\mathrm{s}$ as a result of the fallback accretion, and this timescale is faster than the MHD instability timescale $t_A$. In such a case, the mountain could be sustained by the fallback accretion.
 
\subsection{Ellipticity of mountains}

\begin{figure}
    \centering
    \includegraphics[width=0.47\textwidth]{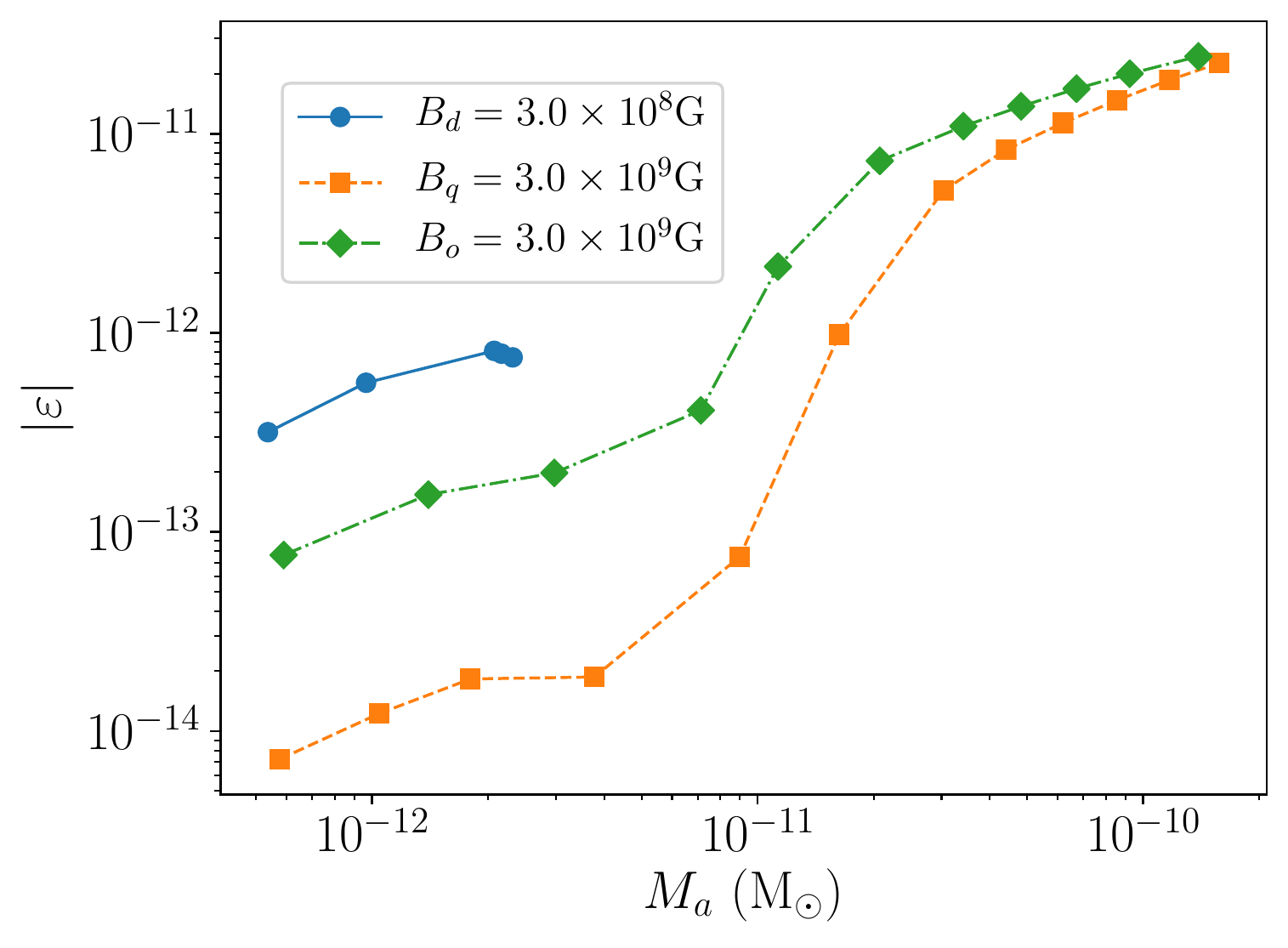}
    \caption{Ellipticitiy as a function of the accreted mass $M_a$. Each point denotes a solution sequence of dipole magnetic field case (blue circles, $B_d = 3.0 \times 10^8~\mathrm{G}$), octupole magentic field case (green diamonds, $B_d = 3.0 \times 10^8~\mathrm{G}$ and $B_o = 3.0 \times 10^9~\mathrm{G}$), and quadrupole magnetic field case (orange square, $B_d = 3.0\times 10^8~\mathrm{G}$ and $B_q = 3.0 \times 10^9~\mathrm{G}$), respectively.}
    \label{fig:epses2}
\end{figure}

Magnetically confined mountains on the neutron star surface are promising candidates for producing continuous gravitational waves. Due to their rapid rotations, millisecond accreting pulsars are the most likely candidates.   

If the strength of the dipole magnetic field is that of typical millisecond pulsars ($3.0 \times 10^{8}~\mathrm{G}$), then the values of ellipticity becomes $|\varepsilon_d| \sim 8 \times 10^{-13}$ (blue circles, $B_d = 3.0 \times 10^8~\mathrm{G}$), $|\varepsilon_o| \sim 2 \times 10^{-11}$ (green diamonds, $B_d = 3.0 \times 10^{8}~\mathrm{G}$ and $B_o = 3.0\times 10^9~\mathrm{G})$ and $|\varepsilon_q| \sim 2 \times 10^{-11}$ (orange square, $B_d = 3.0 \times 10^{8}~\mathrm{G}$ and $B_q = 3.0\times 10^9~\mathrm{G})$ as seen in Fig.~\ref{fig:epses2}. 
The ellipticity of the solutions approximately scales as in equation \eqref{eq:eps}. 
If the neutron star has strong quadrupole and octupole fields, the ellipticity increases by one order of magnitude. The strong multipole magnetic fields support the larger mountains on the surface. 
If the ellipticity for a millisecond pulsar is $|\epsilon| \sim 3 \times 10^{-8}$ \citep[see][]{LIGO_2021arXiv_a}, the necessary value of the magnetic fields is $B_d \sim 3 \times 10^{11}~\mathrm{G}$ (dipole field case) or $B_d \sim 3 \times 10^{10}~\mathrm{G}$ and $B_o$ (or $B_q$) $\sim 3 \times 10^{11}~\mathrm{G}$ (multipole filed case), respectively. The strong multipole magnetic fields can expand the parameter space for the mountain.

The asymmetric magnetic poles and accretion also induce temperature asymmetry on the stellar surface. The asymmetric surface temperature would induce internal temperature asymmetries and increases the ellipticity by generating thermal neutron star mountains (\citealt{Osbore_Jones_2020}). Strong multipole magnetic fields would be important for both magnetically confined mountains and thermal mountains.

\subsection{Hot spots on millisecond pulsars}

Hot spots on millisecond pulsars are created by the accretion of material on to the surface. In the classical model, where a neutron star has only a dipole field, the matter accretes along the dipole field line and falls symmetrically on the north and south poles. The hot spots are equatorially symmetric in this model. As discussed in the Introduction, however, recent observations indicate that the hot spots are not symmetrically located (\citealp{Bilous_et_al_2019,2021ApJ...918L..27R}). If the neutron star's magnetic fields are buried as in solution (q10-b) or solution (o10-b), the hot spots appear asymmetric. Since the neutron star appears to have only a dipole magnetic field, the matter falls along the dipole magnetic field lines (dashed lines in Figs.~\ref{fig:o10-Psi} and ~\ref{fig:q10-psi}). These magnetic field lines, however, do not connect to the north and south poles. As seen in the dotted lines of Figs.~\ref{fig:o10-Psi} and ~\ref{fig:q10-psi}, the matter accretes near $\theta = 80^{\circ}$ and $100^{\circ}$ (solution o10-b) or $\theta = 90^{\circ}$ and $\theta = 180^{\circ}$ (solution q10-b). Therefore, the hot spots naturally become asymmetric or multiple in the same hemisphere if the neutron star has buried higher-order magnetic fields.

\subsection{Fallback accretion and buried magnetic fields}

Central Compact Objects (CCOs) are point sources found at the centres of supernova remnants (\citealp{Enoto_et_al_2019}, for review). Although they appear to be young systems, their period derivatives $\dot{P}$ are too small. This implies that their dipole magnetic fields are weaker than normal pulsars. Nevertheless, some CCOs show magnetar-like activity and are expected to have strong hidden magnetic fields. If certain conditions are satisfied, the magnetic field can be buried by fallback accretion (\citealp{Ho_2011, Vigano_Pons_2012, Zhong_et_al_2021}). 

Suppose a neutron star has strong multipole magnetic fields, and these are buried by accretion, as we described in this paper. In such a case, the multipole magnetic fields are transformed into the negative dipole magnetic field. The neutron star can also sustain the strong toroidal magnetic field within the buried multipole magnetic fields, as seen in Fig.~\ref{fig:Bphi}. As the magnetic fields re-emerge after the fallback (\citealp{Ho_2011,Vigano_Pons_2012}), they would return to the initial state. The toroidal magnetic field is released, and the negative dipole moment is retransformed into the multipole moment. As a result, the dipole magnetic field decreases, and the intense multipole magnetic field appears. The decay of the dipole magnetic field (\citealt{DallOsso_et_al_2012}) could result from the re-emergence of the multipole magnetic fields. 

\section{Summary and conclusion}

We formulated magnetically confined mountains on a neutron star with strong multipole magnetic fields. We confirmed that our new numerical method reproduces well the results of previous works. On a neutron star, as matter accretes on to the magnetic pole, the magnetic fields are buried by the accreted matter. If the neutron star has a dipole magnetic field, the dipole magnetic field is buried and transformed into multipole magnetic fields. We found that the absolute value of the ellipticity increases monotonically up to a certain solution and decreases as in previous works. If the neutron star has a dipole and strong multipole magnetic field, on the other hand, the multiple magnetic fields are buried and transformed into negative dipole magnetic fields. The mass ellipticity of the mountain increases by one order of magnitude if the neutron star has strong multipole magnetic fields. The ellipticity becomes slightly smaller when the mountain has toroidal magnetic fields. If the multipole magnetic fields are buried, they sustain the intense toroidal magnetic field near the stellar surface. The ratio of the toroidal magnetic field to the poloidal magnetic field is close to 100. If the neutron star has buried multipole magnetic fields, the bright spots on the neutron star surface become asymmetric and multiple in the same hemisphere, which aligns with recent observations. The asymmetric surface temperature would induce thermal mountains and increases the ellipticity. Therefore, buried multipole magnetic fields may be the key to better understanding neutron stars. 

More detailed models are needed to theoretically interpret the continuous gravitational waves that will be detected in the future. In this paper, we assumed a polytropic type (isentropic) equation of state and Newtonian gravity. However, the effect of non-isentropic matter and general relativity would be necessary. \citet{Suvrov_Melatos_2019} considered thermal conduction of the magnetically confined mountains and found that GS equilibrium states for polytropic equations of state evolve and reach another equilibrium state over conduction timescales. \citet{Gittins_Andersson_2021} formulated the mountains in general relativity and found that the size of the mountain is suppressed due to the general relativistic effect. The effect of thermal conduction and general relativity would be important, and those are our future works.


\section*{Acknowledgements}

This work was supported by JSPS KAKENHI Grant Number JP20H04728 (KF), JP19K14712, JP21H01078, JP22H01267, JP22K03681 (SK), JP17H06361, JP19K03850 (YK).

\section*{Data Availability}

The data underlying this article will be shared on reasonable request to the corresponding author.




\bibliographystyle{mnras}
\bibliography{bib.bib} 



\newpage

\appendix

\section{Details of the numerical scheme}
\label{App:A}

\begin{figure}
    \centering
    \includegraphics[width=0.47\textwidth]{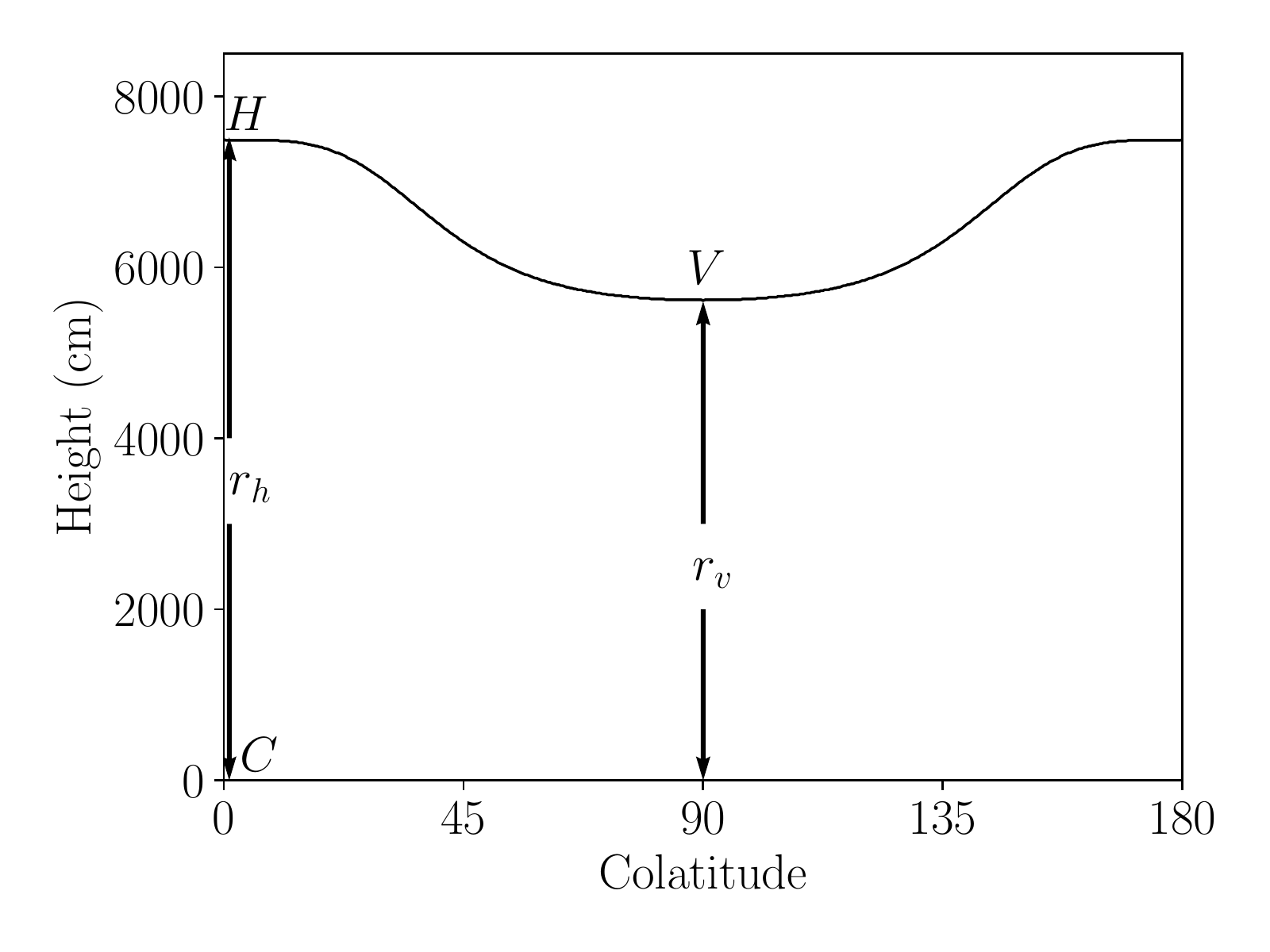}
    \caption{A schematic picture of $r_h$ and $r_v$ of the mountain. The black curve denotes the top of the mountain. $C$, $H$, and $V$ denote the bottom of the mountain, the top of the highest mountain, and the top of the deepest valley, respectively.}
    \label{fig:r_h}
\end{figure}

\begin{figure}
    \centering
    \includegraphics[width=0.47\textwidth]{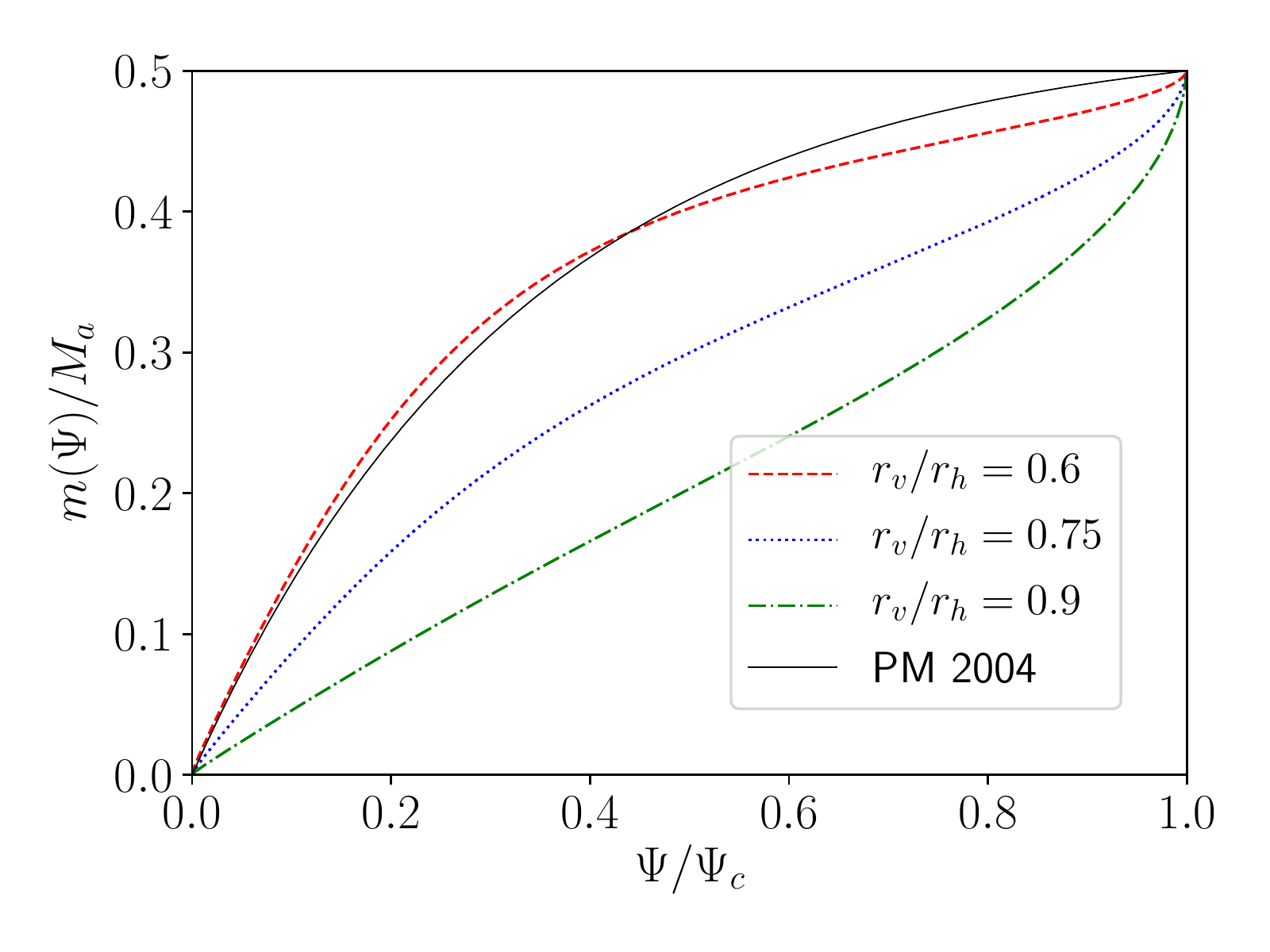}
    \caption{Mass-flux inside the magnetic field lines in the case of the dipole field model. Each line denotes a solution of  our calculations with $r_v/r_h = 0.6$ (red dashed), $r_v/r_h = 0.75$ (blue dotted), and $r_v/r_h = 0.9$ (green dotted-dashed). The black solid line (PM 2004) denotes a mass-flux function in \citet{Payne_Melatos_2004} (equation \ref{Eq:m_Psi}), provided for comparison.}
    \label{fig:M_Psi}
\end{figure}

Our numerical method in this paper is based on a self-consistent field scheme for magnetized equilibrium states. The self-consistent field scheme was initially developed to obtain rotating equilibrium states (\citealp{Ostriker_1964, Hachisu_1986a}). It was then extended for magnetized equilibrium states (\citealp{Tomimura_Eriguchi_2005}). In the self-consistent field scheme, the field equations (e.g., the Poisson equation for self-gravity and the GS equation for magnetic flux) and the matter equation (the first integral of the Euler equation) are calculated iteratively until the calculations converge. Usually, the axis ratio of the equatorial radius to the polar radius is fixed during the iteration to obtain the rotating equilibrium states. 

In our new scheme of magnetically confined mountains, we fix the height of the highest mountain $r_h$ and the deepest valley $r_v$ in Fig.~\ref{fig:r_h} during the iteration. The outer boundary of the computational domain is set above the peak of the highest mountain as $r_{\mathrm{out}} = 1.25 r_h$. Since the density vanishes on the top of the mountains, we obtain the following three equations using equation \eqref{Eq:first_int}. First, for the bottom of the mountain (at C in Fig.~\ref{fig:r_h})
\begin{align}
    K \frac{\Gamma}{\Gamma - 1} \rho_C^{\Gamma-1} = \phi_C + F(\Psi)_C + C,
    \label{Eq:C}
\end{align}
for the top of the highest mountain (at H in Fig.~\ref{fig:r_h})
\begin{align}
    0 = \phi_H + F(\Psi)_H + C,
    \label{Eq:H}
\end{align}
and for the top of the deepest valley between the mountains (at V in Fig.~\ref{fig:r_h})
\begin{align}
    0 = \phi_V + F(\Psi)_V + C.
    \label{Eq:V}
\end{align}
 Here, C, H, and V denote the physical quantities at C, H, and V, respectively. Using these three equations, we obtain the values of $\rho_C$, $C$, and $F_0$. Then, we solve equation \eqref{Eq:first_int} and obtain the density of the mountain $\rho$. Next, we solve equation \eqref{Eq:PsiM} (integrated form of the GS equation) and obtain the magnetic flux function $\Psi$. We iterate these calculations until the system converges. Our numerical scheme is summarized as follows: 
\begin{enumerate}
    \item Fix the functional forms of $F(\Psi)$ and $S(\Psi)$ in equation \eqref{Eq:GSsource}. 
    \item Set initial guess for density and magnetic flux.
    \item Obtain the values of $\rho_C$, $C$ and $F_0$ by solving equations \eqref{Eq:C}, \eqref{Eq:H}, and \eqref{Eq:V}.  
    \item Solve the first integral (equation \ref{Eq:first_int}) and obtain the density.
    \item Solve the integrated form of the GS equation (equation \ref{Eq:PsiM}) and obtain the magnetic flux function.
    \item Iterate from (iii) to (v) until the system converges. 
\end{enumerate}
After obtaining one solution, we increase the value of $r_h$ to increase the accreted mass and calculate the next model in a solution sequence.

As the accreted mass increases, the convergence of the iteration becomes deteriorates. We use an underrelaxation method  (\citealp{Payne_Melatos_2004}; \citealp{Uryu_et_al_2009}) for the magnetic flux function to improve the convergence. If we obtain the provisional value $\Psi_{\mathrm{new}}$ during the iteration, the magnetic flux is updated from the value at the $N$-th iteration cycle $\Psi^{(N)}$ to the $N+1$th $\Psi^{(N+1)}$ as  
\begin{align} 
    \Psi^{(N+1)} = \lambda \Psi_{\mathrm{new}} + (1 - \lambda) \Psi^{(N)},
\end{align}
where $\lambda$ is a softening parameter; we set $\lambda = 0.1$. We use a relative difference as a convergence criteria:
\begin{align}
    \frac{|\Psi^{(N)} - \Psi^{(N-1)}|}{|\Psi^{(N)}|} < \delta,
\end{align}
where we set $\delta = 10^{-6}$. The number of iteration cycles is around 1000 times for one model. 

The accretion pattern depends on the functional form for $F(\Psi)$ and the value of $r_v / r_h$. Previous studies (\citealt{Payne_Melatos_2004,Priymak_et_al_2011,Suvrov_Melatos_2019}) fixed the mass-flux inside the magnetic field lines $m(\Psi)$ instead of the functional form for $F(\Psi)$ in order to realize the magnetic polar accretion. They used the following functional form for $m(\Psi)$:
\begin{align}
    m(\Psi) = \frac{M_a(1 - \exp(-\Psi / \Psi_a))}{2(1 - \exp(-b))},
    \label{Eq:m_Psi}
\end{align}
where $M_a$ is the total accreted mass, $\Psi_a$ is the closed field line inside the accretion disc, and $b = \Psi_c / \Psi_a = 3$ (\citealp{Suvrov_Melatos_2019}). They numerically determined the functional form for $F(\Psi)$ to satisfy equation \eqref{Eq:m_Psi} consistently during the iteration. When the neutron star has strong multipole magnetic fields, however, the accretion process is more complicated (\citealp{Das_et_al_2022}), and it is impossible to fix the mass-flux relation as in equation \eqref{Eq:m_Psi}. Therefore, in this paper, we use the functional form as in equation \eqref{Eq:F_Psi}. The functional form closely reproduces numerical solutions in previous works. Fig.~\ref{fig:M_Psi} shows the mass-flux $m(\Psi)$ of solutions with the functional form in equation \eqref{Eq:F_Psi}. The solid line denotes the mass-flux in \citealp{Payne_Melatos_2004} (equation \ref{Eq:m_Psi}). All three solutions represent the magnetic-polar accretion. The degree of the mass concentration is controlled by changing the ratio $r_v/r_h$ (see dashed, dotted, and dashed-dotted lines in Fig.~\ref{fig:r_h}). As the ratio decreases, the mass concentrates more on the magnetic pole. As seen in Figure~\ref{fig:M_Psi}, the solution with $r_v/r_h = 0.6$ aligns with \citet{Payne_Melatos_2004}. To obtain  solutions with strong multi-pole fields stably, we adopt $r_v/r_h = 0.75$ throughout this paper. Our numerical solutions qualitatively agree with previous works as seen in Sec.~3.1.

\bsp	
\label{lastpage}
\end{document}